\documentclass{aa}  

\usepackage{graphicx}
\usepackage{txfonts}
\usepackage{hyperref}
\usepackage{graphicx}	
\usepackage{amsmath}	
\usepackage{color}
\usepackage{soul}
\usepackage{multirow}
\usepackage{float}
\restylefloat{table}
\usepackage{placeins}

\begin{document} 

  \title{The scarcity of white dwarf-brown dwarf binaries in the solar neighbourhood}
  \subtitle{A population synthesis study}

   \author{A. Santos-García\inst{1}\fnmsep\thanks{E-mail: alejandro.santos.garcia@upc.edu},
          S. Torres\inst{1,2}\thanks{E-mail: santiago.torres@upc.es},
          A. Rebassa-Mansergas\inst{1,2},
          S. L. Casewell \inst{3},
          Zhangliang Chen \inst{4,5},
          Hongwei Ge \inst{6,7,8}
          }

   \institute{Departament de F\'{\i}sica, Universitat Polit\`{e}cnica de Catalunya, c/ Esteve Terrades 5, 08860 Castelldefels, Spain        
         \and
             Institut d'Estudis Espacials de Catalunya, Esteve Terradas, 1, Edifici RDIT, Campus PMT-UPC, 08860 Castelldefels, Barcelona, Spain
         \and
             School of Physics and Astronomy, University of Leicester, University Rd., Leicester LE1 7RH, UK
         \and
             School of Physics and Astronomy, Sun Yat-sen University, Zhuhai 519082, People’s Republic of China
         \and
             CSST Science Center for the Guangdong-HongKong-Macau Great Bay Area, Sun Yat-sen University, Zhuhai 519082
        \and
             Yunnan Observatories, Chinese Academy of Sciences, Kunming, 650216, People’s Republic of China
         \and
             International Centre of Supernovae, Yunnan Key Laboratory, Kunming 650216, People’s Republic of China
         \and
             University of Chinese Academy of Sciences, Beijing 100049, People’s Republic of China
             }

   \date{Accepted XXX. Received YYY; in original form ZZZ}
\titlerunning{The scarcity of WD--BD binaries}
\authorrunning{Santos-García et al.}
 
  \abstract
   {White dwarf–brown dwarf (WD--BD) binaries are intrinsically rare systems that offer a unique opportunity to study binary evolution and the formation of substellar companions. Despite recent observational progress, only a small number of WD--BD systems are currently known in the solar neighbourhood. 
   Understanding whether this scarcity arises from observational biases or instead reflects the underlying physical processes governing the formation and evolution of binaries with substellar companions remains an open question.}
   {We aim to extend binary population synthesis models into the substellar regime and use the resulting simulations to investigate whether the observed scarcity of WD--BD binaries in the solar neighbourhood is consistent with the combined effects of binary formation, binary evolution, and observational selection biases.}
   {We extended the \texttt{MRBIN} binary population synthesis code into the substellar regime by incorporating brown dwarf evolutionary models and an extended initial mass function, enabling the consistent treatment of brown-dwarf companions. We then simulated the stellar population within 100\,pc of the Sun and compared the resulting WD–BD population with the currently known observed sample.}
   {Our simulations predict $13 \pm 5$ WD--BD systems within 100\,pc, in agreement with the observed population. WD--BD binaries thus represent only $\sim 0.1\%$ of the local white dwarf population, confirming their rarity. The models reproduce the overall distributions of white dwarf masses, temperatures, and orbital periods, although the simulated brown dwarf temperatures are systematically higher than those inferred for the observed population. We find that close WD--BD binaries are underproduced when adopting the standard common-envelope efficiency $\alpha_{\mathrm{CE}} = 0.3$, suggesting that larger efficiencies may be required to reproduce the observed short-period systems. The simulations also naturally reproduce the observed deficit of systems at intermediate orbital periods between post-common-envelope binaries and wide non-interacting systems, as well as a weakened brown dwarf desert in main-sequence--brown dwarf binaries.}
   {}


   \maketitle
%



\section{Introduction}
\label{s:intro}

Binary stars provide fundamental constraints on stellar formation and  evolution, since their observed properties preserve information about both their initial conditions and the physical processes governing binary interactions \citep[e.g.,][]{IbenLivio93,Hurley+02,Toonen+13,Toonen+2017}. In particular, binary systems containing white dwarfs (WDs) are key tracers of binary evolution, as they represent the end products of a broad range of evolutionary pathways, including mass transfer episodes and common-envelope (CE) evolution. WD binaries have therefore become powerful laboratories to investigate stellar evolution, angular momentum loss mechanisms, and the physics of compact binary interactions \citep{Nelemans2001a,Zorotovic+10,Schreiber+10,Toonen+12,Zorotovic+22,Torres+25}. However, most previous binary population synthesis (BPS) studies have focused on stellar companions and generally neglected the substellar regime or treated it using simplified prescriptions.

The nature of low-mass companions in binary systems is especially relevant because it probes the transition between stellar and substellar formation regimes. Brown dwarfs (BDs), usually defined as objects with masses below the hydrogen-burning limit ($\sim0.075$--$0.08\,M_{\odot}$), occupy this intermediate regime between giant planets and low-mass stars \citep{Chabrier+97,Burrows+01,Baraffe+03}. Since BDs are unable to sustain stable hydrogen fusion in their interiors, they continuously cool and fade with time, making them intrinsically faint and difficult to detect \citep{Baraffe+03}. Nevertheless, the increasing number of discovered BDs \citep{Kirkpatrick+24}, including systems in binaries, has provided important insights into both stellar and planetary formation processes \citep{Feng+22, Chen+26}.

A well-known characteristic of binary populations is the so-called BD desert, namely the observed lack of BD companions orbiting solar-type stars at short orbital periods \citep{Halbwachs+00,Marcy+00,Grether+06,Troup+16,Stevenson+23}. Radial velocity and transit surveys have shown that the frequency of BD companions is significantly lower than that of both planetary and stellar companions at short orbital periods. The origin of this deficit remains uncertain, although several explanations have been proposed, including different formation channels for low- and high-mass BDs \citep{Chabrier+03,MaGe14}, dependencies on the initial mass-ratio distribution \citep{Reggiani+13}, and the effects of binary evolution and observational biases \citep{Wallace+26}. However, despite numerous attempts to explain the BD desert, the currently known sample remains too limited to provide conclusive evidence in favour of any particular scenario \citep{Vowell+25,Giacalone+26}.

BDs orbiting WDs provide a particularly interesting framework to study both substellar formation and binary evolution as well as the BD dessert. During the evolution of the primary star, physical processes such as stellar mass loss, stellar winds, stable mass transfer and CE evolution can strongly modify the orbital properties of the system \citep{Nordhaus+10,Nordhaus+13}. In close binaries, unstable mass transfer may trigger a CE phase \citep{Pac76,Webbink08}, during which the BD spirals in within the envelope of the giant star. Systems surviving this process can emerge as compact  white dwarf–brown dwarf binaries (WD–BD binaries) with orbital periods of only a few hours \citep{Maxted+06,Casewel+18,Parsons+17-2,Parsons+2025}. However, many systems are expected to merge during the CE phase, especially those with extreme mass ratios \citep{Zorotovic+22,Chen+24}.

Despite recent observational progress, only a small number of WD--BD binaries are currently known. \citep{Amaro+23, French+25, Chen+26}. Wide WD--BD binaries have also been identified through infrared imaging and microlensing surveys \citep{Becklin+88,Luhman+11, French+23,Zhang+24}. However, the observed population remains remarkably small compared to the large number of known MS--BD binaries \citep{Gaia23,Stevenson+23,Chen+26}. This scarcity may reflect both observational limitations, due to the intrinsic faintness of BDs, and the strong impact of binary evolutionary processes on their survival probability. In particular, as already stated, systems with extreme mass ratios are expected to be strongly affected by interaction processes such as CE evolution, which may significantly reduce their survival probability. 

A consistent theoretical framework capable of modelling both stellar and substellar companions is therefore required in order to interpret the observed population of WD--BD systems and to assess the relative importance of formation and evolutionary effects. Extending BPS models into the substellar regime requires both a realistic description of the initial mass function (IMF) at very low masses and dedicated BD evolutionary models. Observational and theoretical studies suggest that the IMF changes slope in the substellar regime \citep{Kroupa01,Chabrier+03,Kroupa+13, Sollima19}, while the thermal evolution of BDs must be described using specialised cooling tracks \citep{Baraffe+03, BHAC15}. Incorporating these ingredients into BPS simulations is therefore essential to consistently model binaries containing substellar companions. 

In this work, we present an updated version of the \texttt{MRBIN} BPS code that extends for the first time a rigorous treatment of companions down to the BD regime. By incorporating BD evolutionary sequences and an extended substellar IMF, we simulate the stellar population within 100\,pc of the Sun and investigate the expected population of WD--BD binaries. Our goal is to determine whether the observed scarcity of these systems can be naturally reproduced by current binary evolution models, and to evaluate the role played by CE evolution in shaping their orbital and physical properties.

\section{The population synthesis modelling}
\label{s:model}

\subsection{The Monte Carlo simulator}
\label{ss:MonteCarlo}

To simulate the stellar population of the Solar neighbourhood, including both single and binary stars, we make use of an updated version of \texttt{MRBIN}, a population synthesis code developed by our research group and extensively characterised in previous studies \citep{Torres+98,GarciaBerro+99,Camacho+14,Cojocaru+17,Santos+25}. The code is capable of reproducing the observed properties of single main-sequence stars and WDs \citep[e.g.][]{Torres+01,Torres+19}, and now also includes the BD regime. In addition, \texttt{MRBIN} can model binary populations \citep[e.g.][]{Torres+22,Santos+25}, whose stellar evolution is computed using the Binary Stellar Evolution (\texttt{BSE}) code developed by \citet{Hurley+02}, including several updates specifically focused on WD binaries \citep[e.g.][]{Camacho+14,Cojocaru+17,Canals+18, Santos+25}.

In this work, we adopt the Galactic model described in \citet{Santos+25} and references therein to simulate the stellar population within 100\,pc of the Sun. This model includes the different Galactic components through their corresponding age, metallicity, and spatial distributions.

For the binary population, we follow the prescriptions adopted by \citet{Torres+22}, which provide the best fit to the observed {\it Gaia} white dwarf binary population within 100 pc after exploring a wide range of initial binary population parameters. Specifically, we assume a binary fraction of 32\% together with a mass-ratio distribution of the form $n(q)\propto q^{-1.13}$, where $q=M_{2}/M_{1}$ and $M_{2}$ correspond to the mass of the secondary component. We also adopt a thermal eccentricity distribution \citep{Heggie75} and an orbital separation distribution given by $f(a)\propto a^{-1}$.

The evolution of binary systems is treated through a CE formalism characterised by the parameter $\alpha_{\rm CE}$ \citep{Tout+97,Webbink08}, which quantifies the efficiency with which orbital energy is transferred to the envelope of the donor star during the CE phase, eventually leading to its ejection. This process is particularly important for close WD--BD binaries, many of which are expected to merge during the CE phase rather than survive as detached systems. In our reference model we adopt $\alpha_{\rm CE}=0.3$ \citep{Camacho+14,Santos+25}, although, as discussed later in Section~\ref{ss:ce}, larger efficiencies may be required to reproduce the observed population of close WD--BD systems. The envelope binding energy in \texttt{MRBIN}, an updated version of BSE \citep{Hurley+02}, is calculated following the prescription of \citet{Claeys+14}, with the effective envelope binding-energy parameter $\lambda$ determined from the evolutionary properties of the donor star.

The population synthesis procedure begins by generating an initial distribution of single and binary stars according to the prescriptions described above. These systems are subsequently evolved to the present time in order to obtain the expected present-day stellar populations. Since \texttt{MRBIN} is particularly focused on WDs and WD binaries, the code incorporates the latest evolutionary sequences computed by the La Plata group \citep[e.g.][]{Camisassa+16,Camisassa+19}. From the resulting WD masses and cooling ages, these models provide the corresponding stellar parameters, including effective temperatures, radii, luminosities, and surface gravities. 

The latest version of \texttt{MRBIN} now extends into the BD regime through the incorporation of low-mass star and BD evolutionary models from \citet{BHAC15} and \citet{Phillips+20}. In a similar way to the WD treatment, the masses and ages of low-mass stars and BDs are used to derive their expected stellar parameters. These models are applied to objects with masses below $0.11\,M_{\odot}$, after exploring a wide range of transition masses to ensure a smooth overlap between the previous low-mass limit of the simulations and the new extension into the substellar regime.

The inclusion of BDs also requires extending the adopted initial mass function into the substellar domain. The prescription used to generate the initial mass distribution of low-mass stars and BDs is described in detail in Section~\ref{ss:imf}.

Synthetic photometry is generated for all objects in order to directly compare the simulations with observations. For WDs, the evolutionary sequences from the La Plata group provide the fundamental stellar parameters, including mass, cooling age, effective temperature, luminosity, and surface gravity. These parameters are then combined with the hydrogen-rich atmosphere models of \citet{Koester+10} to derive synthetic magnitudes in the \textit{Gaia} $G$, $BP$, and $RP$ passbands, which are used throughout this work for the comparison with the observational data.

For low-mass stars and BDs, the \citet{BHAC15} models also provide synthetic \textit{Gaia} photometry. Main-sequence magnitudes are instead derived through a three-dimensional spline interpolation in luminosity, effective temperature, and metallicity using the PARSEC evolutionary tracks \citep{Bressan+12}, following the procedure validated in \citet{Santos+25}. In that work, the simulations were limited to absolute \textit{Gaia} magnitudes brighter than $M_{G}\sim14.2$\,mag due to the absence of low-mass star and BD models. The inclusion of the new substellar evolutionary sequences now extends this limit down to $M_{G}\sim19$\,mag.

Following the observational capabilities of \textit{Gaia}, synthetic binaries are classified as either resolved or unresolved systems. A binary is considered unresolved when its angular separation is smaller than 2\arcsec \,\citep{Torres+22}, in which case \textit{Gaia} detects the system as a single source rather than as two independent stellar objects. This threshold is more conservative that the nominal \textit{Gaia} angular resolution of  $\sim0.5$\arcsec\ \citep{Gaia16}, which is intended to account for deblending effects and systems with similar component brightnesses. The \texttt{MRBIN} code computes the projected angular separation of each system from its orbital separation and parallax, taking into account the orbital eccentricity, simulated inclination, and orbital phase. As a consequence, some binaries with large orbital periods may still appear unresolved. For unresolved systems, the magnitudes are computed by combining the flux contributions from both stellar components. 

Finally, in order to reproduce the observational characteristics of the \textit{Gaia} catalogue as realistically as possible, synthetic photometric and astrometric uncertainties are incorporated following \citet{Riello+21}\footnote{\url{https://www.cosmos.esa.int/web/gaia/science-performance}}. We also assign a synthetic \textit{phot\_bp\_rp\_excess\_factor} parameter \citep{Riello+21} to each object to account for flux contamination from nearby sources. It is important to note that the synthetic photometric uncertainties strongly depend on the magnitude of the objects. As a result, sources brighter than $\sim13$\,mag exhibit only a small dispersion in BP after the errors are introduced, whereas objects fainter than $\sim14$--15\,mag show a dramatic increase in the BP dispersion. This effect becomes particularly severe towards fainter magnitudes and is especially relevant for the BD population, which occupies precisely this low-luminosity region of the {\it Gaia} Hertzsprung-Russell diagram (HRD). The implications of these photometric effects for the WD--BD population are discussed in Section~\ref{ss:simwdbd}.

\subsection{The initial mass function in the low-mass and substellar regimes}
\label{ss:imf}

As discussed in the previous section, we have updated our population synthesis code by incorporating low-mass star and BD evolutionary models. This extension allows our models to reach regions of the \textit{Gaia} HRD that were previously inaccessible, while providing a more complete representation of the stellar and substellar populations of the Solar neighbourhood. Previous version of the code were limited to an initial mass range between 0.09 and 50\,$M_{\odot}$. The updated version extends this lower limit down to 0.01\,$M_{\odot}$, entering the BD regime. Consequently, an extension of the adopted initial mass function (IMF) into the substellar regime is required in order to consistently describe both stellar and substellar populations over the entire mass range from 0.01 to 50\,$M_{\odot}$.

To reproduce the initial mass distribution of single stars, as well as the primary masses of binary systems, we adopt the IMF of \citet{Kroupa01} in the stellar regime. This prescription assumes a slope of $-1.3$ for masses between 0.16 and 0.5\,$M_{\odot}$ and a slope of $-2.3$ above 0.5\,$M_{\odot}$. Towards lower stellar masses, we extend the IMF using the prescription of \citet{Sollima19}, adopting a slope of $+1.3$ in the interval between 0.08 and 0.16\,$M_{\odot}$.

\begin{figure} [h!]
   \centering
   \includegraphics[width=1\hsize]{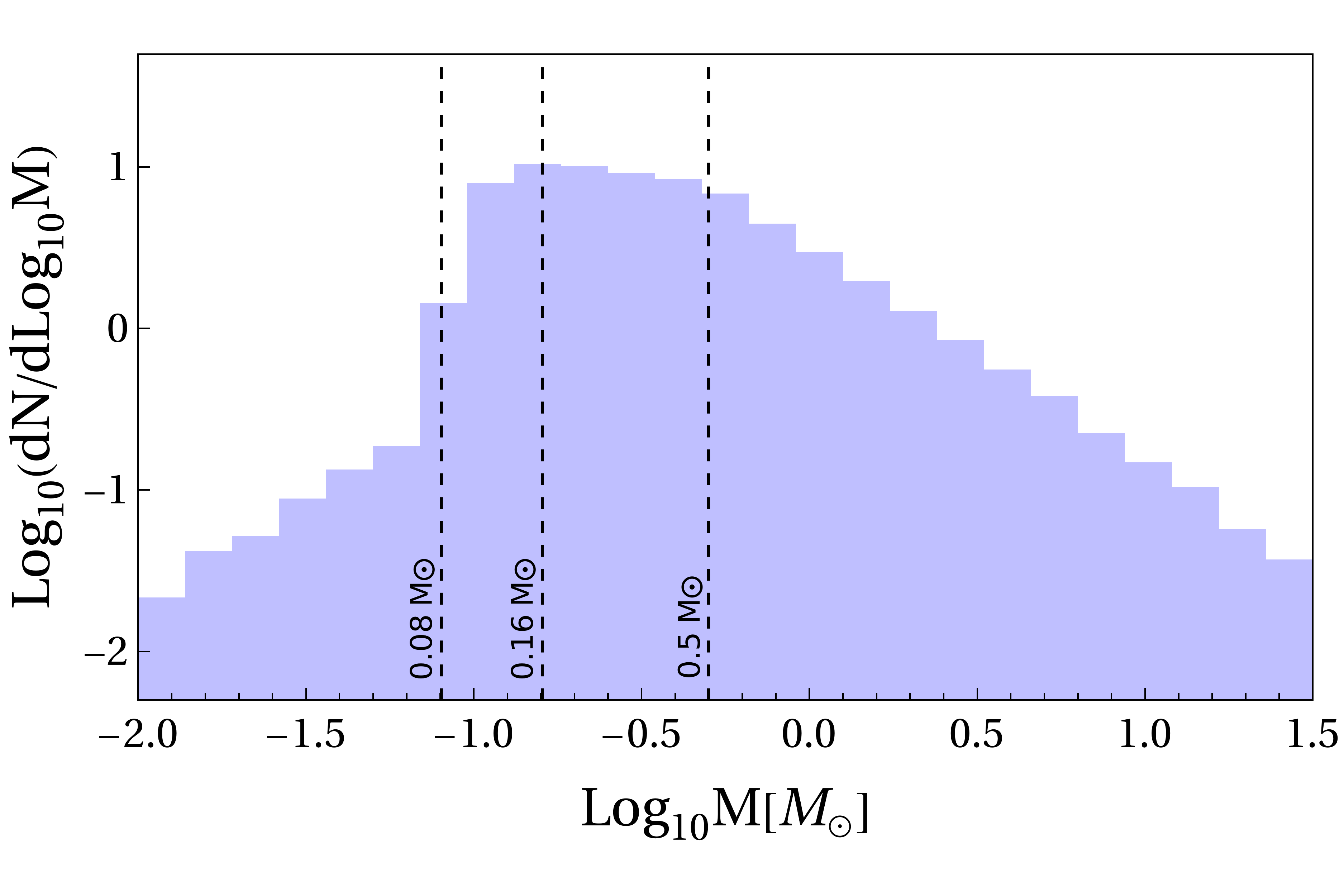}
   \caption{Simulated initial mass distribution resulting from the extended IMF adopted in this work. Dashed lines show the mass limits in which the slope of the theoretical IMF changes. 
   } 
   \label{f:imf}
\end{figure}

However, extending the same \citet{Sollima19} IMF continuously into the BD regime leads to unrealistic substellar populations, producing an underrepresentation of very low-mass BDs while simultaneously generating an excess of objects close to the hydrogen-burning limit around $0.08\,M_{\odot}$. This suggests that a single continuous IMF is not sufficient to accurately reproduce both stellar and substellar populations. We therefore introduce a separate canonical BD IMF with a shallower slope of $0.3$ below $0.08\,M_{\odot}$, following the framework proposed by \citet{Kroupa+13}, where stellar and substellar IMFs partially overlap. Thus, the IMF adopted in this work follows this structure:

\begin{equation}
\xi(M) \propto
\begin{cases}
k\,M^{0.3}, & M < 0.08\,M_{\odot} \\
M^{1.3}, & 0.08 \leq M < 0.16\,M_{\odot} \\
M^{-1.3}, & 0.16 \leq M < 0.5\,M_{\odot} \\
M^{-2.3}, & M \geq 0.5\,M_{\odot}
\end{cases}
\end{equation}

Figure~\ref{f:imf} shows the resulting initial mass distribution, $\xi(M)$, obtained when applying the IMF prescription described above within a representative simulation generated with \texttt{MRBIN}.

\begin{figure} [h!]
   \centering
   \includegraphics[width=0.9\hsize]{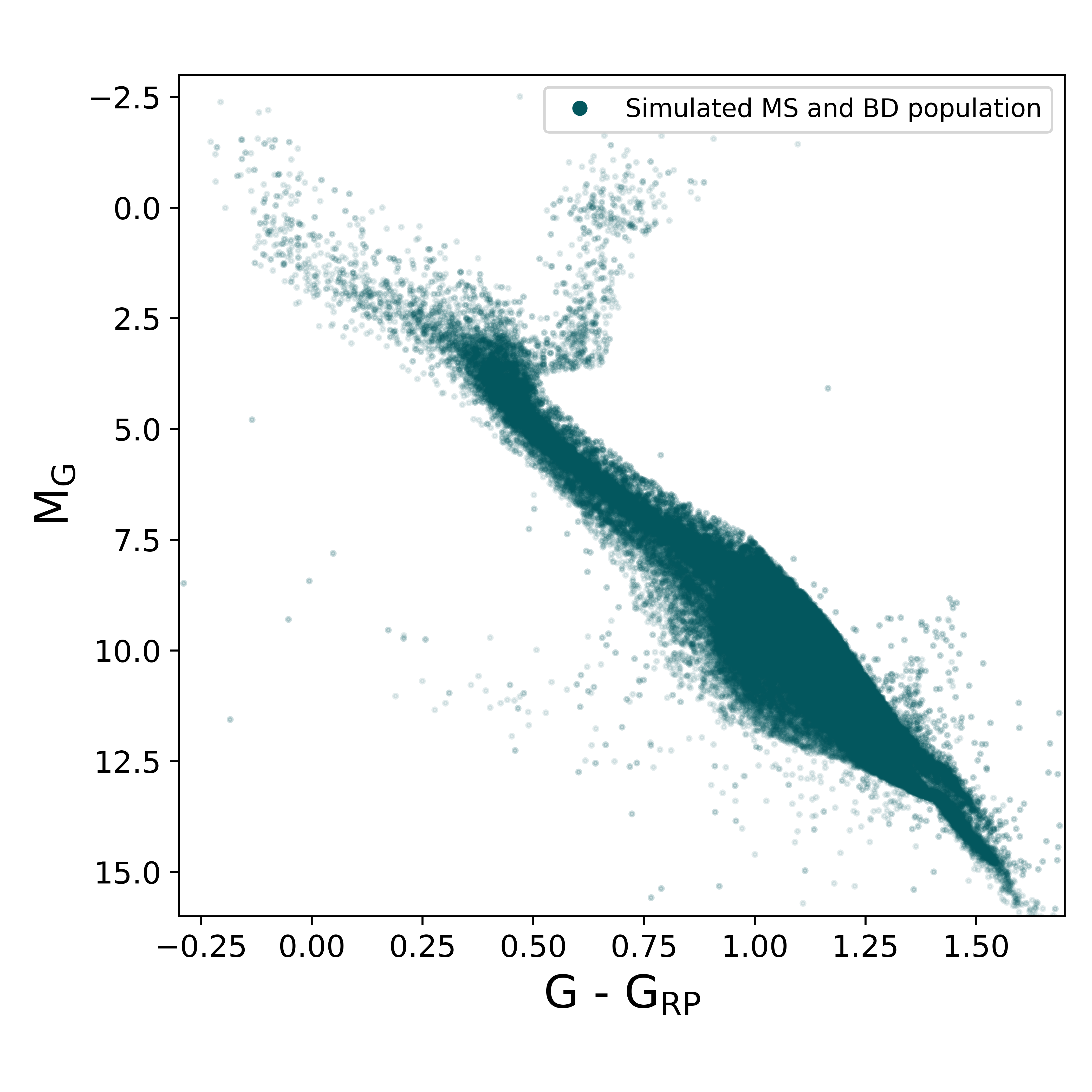}
   \includegraphics[width=0.9\hsize]{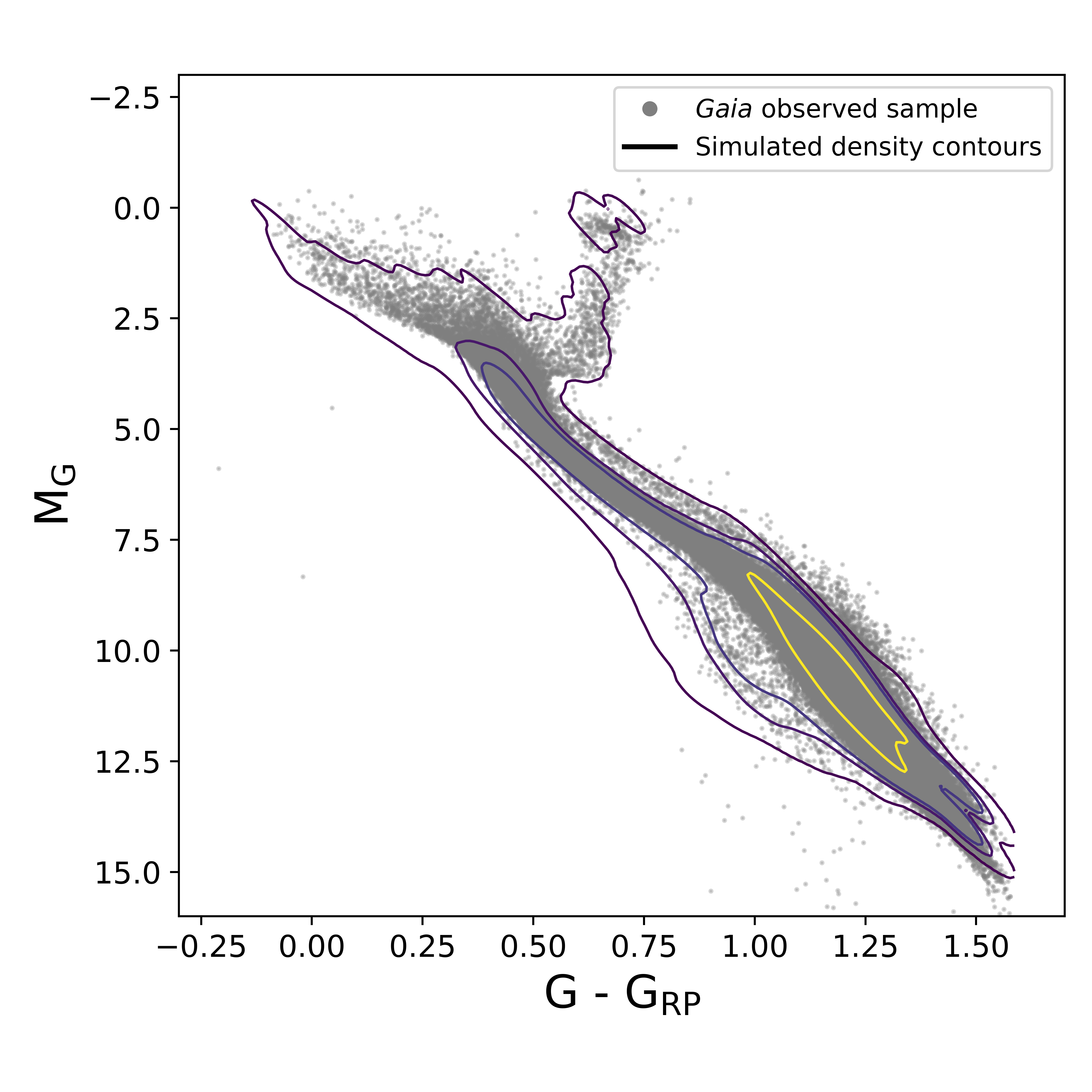}
   \caption{Top panel: Simulated {\it Gaia} HRD of the main-sequence and BD population within 100\,pc obtained with the updated \texttt{MRBIN} code. 
Bottom panel: Comparison between the observed {\it Gaia} HRD within 100\,pc (gray dots) and the synthetic stellar population generated with the updated \texttt{MRBIN} code (contour lines). The extension into the substellar regime allows the simulated HRD to reproduce the BD population and to reach lower luminosities and cooler effective temperatures.}
      \label{F:simdenmap}
   \end{figure}

Note that, we introduce a normalisation factor, $k$, in the BD regime in order to account for the lower formation efficiency of substellar objects relative to stars. Several studies have suggested that BDs may not follow the same continuous IMF as stars, but instead constitute a partially distinct population with different formation channels and pairing properties \citep[e.g.][]{Thies+07,Kroupa+13}. While \citet{Kroupa+13} adopted values of $k\sim1/3$, we use a lower normalisation of $k=0.1$. This choice is motivated by the comparison between the synthetic and observed \textit{Gaia} HRDs (see Fig. \ref{F:simdenmap}), where a larger $k$ leads to a significant overproduction of objects in the low-mass and substellar regions, particularly BD--BD systems. A value of $k=0.1$ provides a substantially better agreement with the observed density of sources in this part of the diagram.

To further validate the adopted normalisation of the substellar IMF, we compared the simulated BD population within 20\,pc with recent observational constraints on the local volume-limited census \citep[e.g.][]{Kirkpatrick+24}. We tested two normalisations, $k=0.1$ and $k=1/3$. The observed 20\,pc census implies a BD-to-stellar ratio of approximately $1:4$, whereas our simulations yield ratios of $\sim 1:4.5$ for $k=0.1$ and $\sim 1:2.4$ for $k=1/3$. The $k=1/3$ model therefore overproduces BDs within the local volume, while the $k=0.1$ model is in much better agreement with the observations. We note, however, that even in the $k=0.1$ case the total number of objects predicted within 20\,pc ($\sim 4500$) remains moderately higher than the observed census, which is expected given that the current observational sample is still affected by incompleteness at the lowest masses and in unresolved multiple systems. We therefore adopt $k=0.1$ as the fiducial normalisation of the substellar IMF throughout this work.

Finally, it is important to note that the WD--BD population studied in this work is barely affected by the exact value of $k$. In most WD--BD progenitors, the primary star evolves into a WD and is therefore drawn from the stellar IMF, while the BD companion is assigned through the adopted mass-ratio distribution. Consequently, variations in the normalisation of the BD IMF primarily modify the number of isolated BDs and BD--BD systems, while leaving the predicted WD--BD population unchanged.

\section{The observed sample of white dwarf - brown dwarf binaries within 100\,pc}
\label{s:obs}

\begin{table*}[h!]
\centering
\caption{Known white dwarf--brown dwarf binary systems within 100\,pc from the Sun used for comparison with our population synthesis models.\\
Ref: \textbf{1.} \citet{Mace+13}; \textbf{2.} \citet{Mace+18}; \textbf{3.} \citet{Burgasser+25}; \textbf{4.} \citet{Luhman+11}; \textbf{5.} \citet{Rodriguez+11}; \textbf{6.} \citet{Giammichele+12}; \textbf{7.} \citet{Legget+17}; \textbf{8.} \citet{Meisner+20}; \textbf{9. }\citet{Zhang+20}; \textbf{10.} \citet{Giammichele+16}; \textbf{11.} \citet{Filippazzo+15}; \textbf{12.} \citet{Becklin+88}; \textbf{13.} \citet{Kirkpatrick+93}; \textbf{14.} \citet{Kirkpatrick+99}; \textbf{15.} \citet{Bedard+17}; \textbf{16. }\citet{Gonzales+22}; \textbf{17.} \citet{Bravo+25}; \textbf{18.} \citet{Steele+09}; \textbf{19.} \citet{Casewell+24};\textbf{ 20. }\citet{Day-Jones+11}; \textbf{21.} \citet{Baig+24}; \textbf{22.} \citet{Zhang+242}; \textbf{23.} \citet{Deacon+14}; \textbf{24.} \citet{Best+24}; \textbf{25.} \citet{Vincent+24}; \textbf{26.} \citet{Farihi+04}; \textbf{27.} \citet{Farihi+05}; \textbf{28.} \citet{Jimenez+18}; \textbf{29.} \citet{Bergeron+21}; \textbf{30.} \citet{Casewell+24-2}; \textbf{31.} \citet{Amaro+23}; \textbf{32.} \citet{Steele+13}; \textbf{33.} \citet{Burleigh+06}; \textbf{34.} \citet{Maxted+06}.\\
Distances obtained from \citet{Gaia23}.}
\label{Tab1}
\begin{tabular}{l|ccccccc}
\hline
\hline
\textbf{Name} &$T_{\rm eff,WD}$ (K) & $T_{\rm eff,BD}$ (K) & $M_{\rm WD}$ ($M_\odot$) & $M_{\rm BD}$ ($M_\odot$)  & Dist (pc) & Period (days) & Ref\\
\hline
\hline
WOLF 1130         & $<7000$& 621 $\pm$ 9 & 1.24$^{+0.19}_{-0.15}$ & 0.043$^{+0.006}_{-0.005}$  & 16.58  & 56851893    & 1-3\\
WD 0806-661       & 10205  & 330-350   & 0.58  $\pm$ 0.03 & 0.007-0.009  & 19.23  & 57849756   & 4-7\\
LSPM J0055+5948   & 4734   & 800 $\pm$ 82  & 0.47 $\pm$ 0.01  & 0.053 $\pm$ 0.009  & 22.84  & 4056357   & 8\\
COCONUTS-1        & 5115   & 1255$^{+6}_{-8}$       & 0.55 $\pm$ 0.02 & 0.066$^{+0.002}_{-0.003}$  & 31.47  & 21333011   & 9\\
GD 165AB          & 12392  & 1755 $\pm$ 102 & 0.64 $\pm$ 0.02  & 0.060$\pm$0.015  & 33.40  & 594773    &10-16\\
LPSM J0806+2215   & 5702   &1841       & 0.56  & 0.076  & 49.43  & 3473871 &17\\
PHL 5038AB        & 7525   & 1400-1500 & 0.72 $\pm$ 0.15  & 0.069 $\pm$ 0.001  & 73.45  & 255640    & 18,19\\
LSPM 1459+0857    & 5436   & 1300-1500 & 0.59  & 0.066 $\pm$ 0.006 & 74.24  & 1421648040 & 20\\
VVV J1256-62AB    & 4440   & 2298$^{+45}_{-43}$ & 0.62 $\pm$ 0.04 & 0.082 $\pm$ 0.001  & 75.59  & 22107382     & 21,22\\
LSPM J0241+2553   & 6620   &1874 $\pm$ 201      & 0.64  &  0.050$\pm$ 0.026     & 76.29   & 51014994 &23-25\\
\hline
GD 1400           & 11386  & $\sim$1650      & 0.68 $\pm$ 0.03  & 0.074 $\pm$ 0.006  & 46.25  & 0.41583    &26-30\\
NLTT 5306         & 7756   & 1500-1600 & 0.44 $\pm$ 0.04  & 0.050 $\pm$ 0.003  & 79.17  & 0.07076    &31,32\\
WD 0137$-$349     & 16500  & 1300-1400      & 0.39 $\pm$ 0.04  & 0.053 $\pm$ 0.006 & 101.55 & 0.08028   &33,34\\
\hline
\end{tabular}
\end{table*}

BDs, and especially WD--BD binaries, are observationally challenging systems due to the intrinsically low luminosities and cool temperatures of BDs, which make them difficult to detect, particularly at optical wavelengths. The complete compilation of observed WD--BD systems contains a total of 21 objects, the most distant of which is KMT-2020-BLG-0414, located at 1330 pc \citep{Zhang+24}. 
In this work we compiled a sample of 13 WD--BD binaries located within 100\,pc of the Sun. From these 13 systems, 3 of them are short period post-common-envelope binaries (PCEBs) while 10 of them are wide binaries. The observational properties for these systems were collected from several studies in the literature, and the parameters relevant to this work are presented in Table~\ref{Tab1}.

One particular object, WOLF~1130, deserves special consideration. Although it is commonly included in catalogues of WD--BD binaries (e.g. Chen et al. 2026), it is in fact a triple system composed of an M subdwarf star (WOLF~1130A), a WD (WOLF~1130B), and a BD companion (WOLF~1130C). In this system, WOLF~1130AB forms a close tidally bound binary, in which the M subdwarf is expected to eventually transfer mass onto the WD companion, first evolving into a cataclysmic variable and later merging. The BD, WOLF~1130C, is instead a wide outer companion orbiting the central WOLF~1130AB pair (Mace et al. 2018).

Among the systems in our sample, two objects (LSPM J0241+2553AB and LSPM J0806+2215AB) lacked measurements of both the orbital period and the WD mass. The WD components in both systems are classified as DA WDs \citep{Limoges+15, Garciazamora+23}. We therefore interpolated their $Gaia$ magnitudes in the cooling sequence of La Plata to derive the photometric WD masses, which are listed in Table\,\ref{Tab1}. Furthermore, although not included in the table, we found an orbital separation of $385 \pm 2.5$ AU \citep{Bravo+25} for LSPM J0806+2215AB and $2380\pm 36$ AU \citep{Best+24} for LSPM J0241+2553AB. Using the estimated masses of both components, we also calculated the corresponding orbital periods, which are reported in Table\,\ref{Tab1}. It is worth noting that the BD mass estimate for LSPM J0241+2553AB carries a large uncertainty, which could significantly affect the calculated orbital period.

In the case of GD 165AB, no effective temperature measurement for the BD companion has been reported, although the remaining system parameters are sufficiently well constrained for the system to be included in our analysis. A similar situation occurs for VVV J1256-6202, for which no BD $T_\mathrm{eff}$ estimate has been found.

Finally, WD 0137-349 has been included in the sample despite being located slightly beyond our nominal distance limit, at 101.55\,pc according to \citet{Gaia16}. Given its very small offset from the adopted 100\,pc boundary, we consider it consistent with the local volume explored in this work and therefore retain it in the observational sample.

The $Gaia$ photometry of the observed systems is listed in Table \ref{Tab2}. Several important caveats arise from the observational sample. WOLF~1130B is not independently detected by $Gaia$, and the measured photometry likely corresponds to the combined flux of the M subdwarf and WD components. The system is located within the unresolved WD-MS region defined by Rebassa-Mansergas et al.~(2021), whose synthetic population was later studied by Santos-Garc\'{\i}a et al.~(2025). The BD companion in this system also remains undetected.

Two additional systems, PHL~5038AB and NLTT~5306, are identified as double stars in $Gaia$. Their measured magnitudes therefore correspond to the combined flux of both components, and these systems can effectively be considered unresolved $Gaia$ binaries. In contrast, VVV~J1256$-$62AB is the only system in the sample for which $Gaia$ resolves both the WD and BD components individually. Nevertheless, the BD companion exhibits an absolute magnitude fainter than 20\,mag and an anomalous BP$-$RP colour.

For the remaining nine systems only the WD companion is detected by $Gaia$, with no nearby objects sharing compatible proper motions that could correspond to resolved BD companions. Some of these objects appear in the canonical WD sequence region in the HRD where also the unresolved systems are expected to be, which could be naturally explained if they correspond to unresolved WD--BD binaries, in agreement with our simulations.

\section{Analysis of the white dwarf--brown dwarf population}
\label{s:pop} 

Following the methodology described in \citet{Santos+25}, we generated synthetic stellar and binary populations within 100\,pc of the Sun. As in that work, the simulations were normalised to reproduce the observed WD population within this volume, resulting in a sample containing approximately 12,000 WDs. The updated version of \texttt{MRBIN}, which now incorporates low-mass stars and brown dwarfs together with the extended IMF described in Section~\ref{ss:imf}, allows us to model substellar populations and binary systems involving BDs, including BD--BD, MS--BD, and WD--BD binaries.

The resulting population is shown in Figure~\ref{F:simdenmap}. The synthetic HRD reproduces the characteristic morphology of the low-mass stellar population, including the two distinct sequences associated with single stars and binary systems at magnitudes 12-15. The bottom panel compares the observed \textit{Gaia} 100\,pc sample with the simulated density distribution. Overall, the agreement is good, particularly at the faint end of the main sequence where both sequences are clearly reproduced, although a small offset between the simulated and observed samples remains. In the following sections, we focus on the properties of the simulated WD--BD population.

\subsection{The simulated WD--BD population at 100\,pc}
\label{ss:simwdbd}
\begin{figure}[h!]
   \centering
   \includegraphics[width=0.9\hsize]{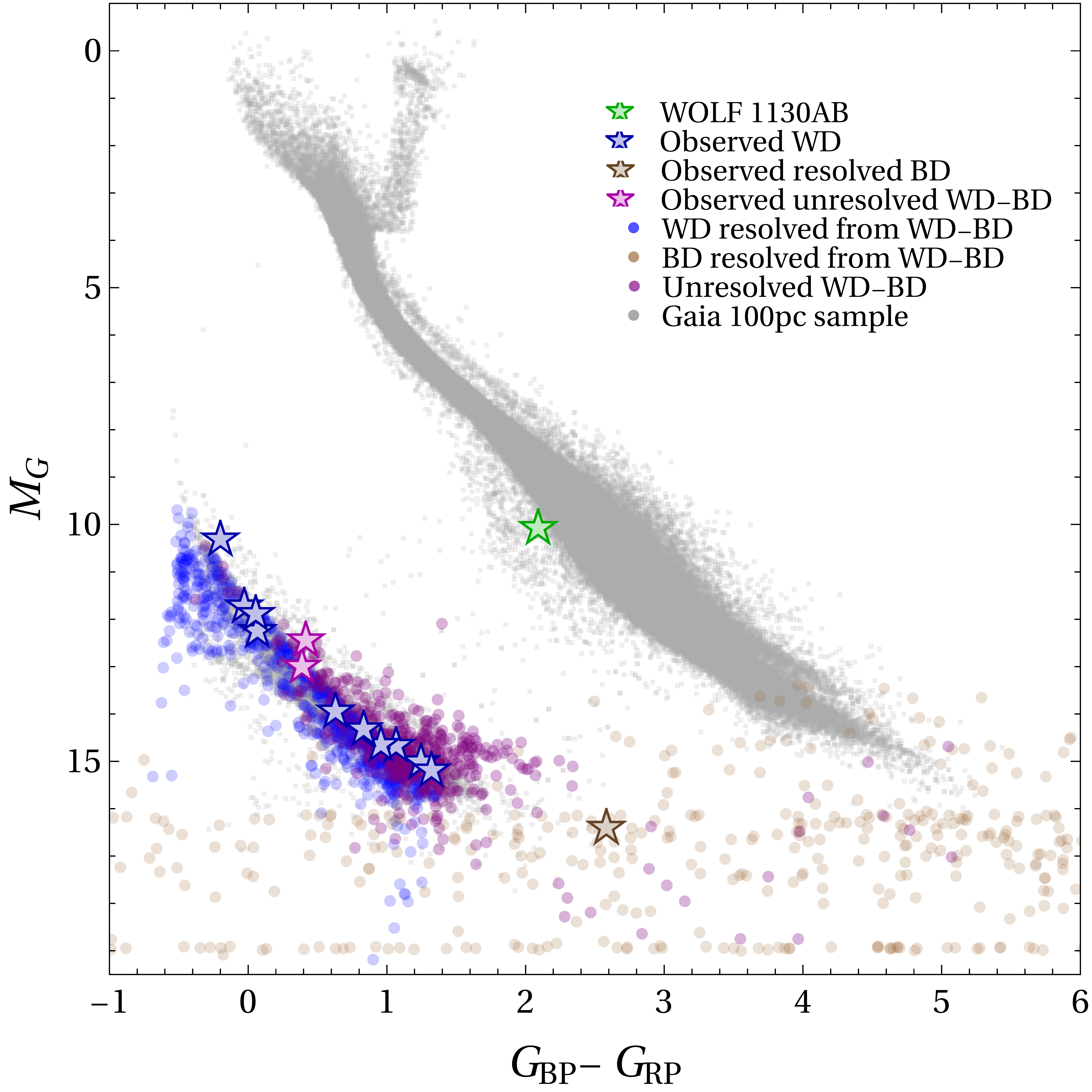}
   \includegraphics[width=0.9\hsize]{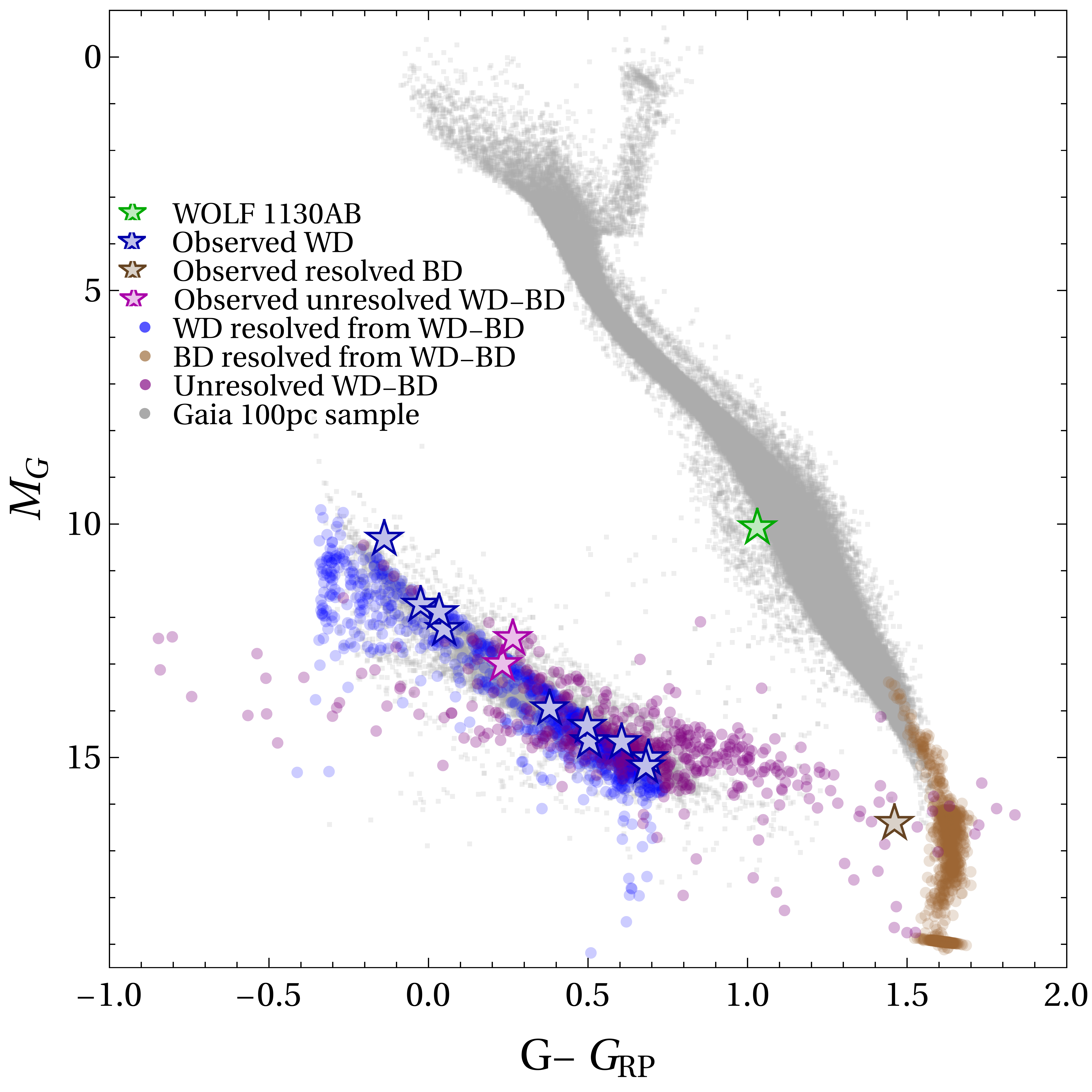}
   \caption{$Gaia$ HRDs of the simulated stellar population within 100\,pc together with the observed WD--BD systems. Top panel: $M_G$ versus $BP-RP$. Bottom panel: $M_G$ versus $G-RP$. Purple points indicate unresolved simulated WD--BD binaries, brown points represent the BD companions for resolved systems while blue points are the WD companion. The different observed systems are shown as stars. The simulations reveal the expected location of unresolved WD--BD systems and illustrate the larger photometric uncertainties affecting the BP band at faint magnitudes.}
    \label{Fighrd}
   \end{figure}

We simulated the stellar population of the solar neighbourhood with particular emphasis on the predicted population of WD--BD binaries and on how different model prescriptions affect their properties. Figure~\ref{Fighrd} presents the distribution of the synthetic population in the $Gaia$ HRD together with the observed sample. To improve the statistical representation of the simulated WD--BD systems in the diagrams, we combined the results from more than 100 independent Monte Carlo realisations.

The {\it Gaia} HRD in the $M_G$-$(G_{\rm BP}-G_{\rm RP})$ plane (top panel) illustrates how the synthetic photometric uncertainties incorporated into our simulations strongly affect the BP photometry at magnitudes fainter than $G \sim 13$-14\,mag. No photometric selection cuts were applied in this case in order to highlight the strong sensitivity of BDs to these BP-related issues. Applying flux cuts or observational selection criteria would drastically reduce the number of detectable WD--BD systems, which is particularly problematic given the intrinsically small size of the population.

In contrast, the {\it Gaia} HRD in the $M_G$--$(G-G_{\rm RP})$ plane (bottom panel) is largely unaffected by these effects. Because very cool and low-luminosity objects emit only a small fraction of their flux in the BP band, the corresponding $G_{\rm BP}$ measurements become increasingly uncertain towards the faint end of the sequence, whereas the $G_{\rm RP}$ photometry remains comparatively robust. This suggests that searches for WD--BD systems based on {\it Gaia} photometry may benefit from using the $G-G_{\rm RP}$ colour rather than $G_{\rm BP}-G_{\rm RP}$, particularly in the low-luminosity regime where BP measurements become increasingly uncertain.

The {\it Gaia} HRDs of Fig.\,\ref{Fighrd} also reveal a relatively well-defined region where unresolved WD--BD systems, shown in purple, are expected to lie. 
Targeted searches within this region may help uncover previously unidentified WD--BD binaries.
However, this task remains challenging because the same area of the {\it Gaia} HRD is heavily contaminated by sources with poor astrometric solutions, high values of the photometric excess factor due to chance alignments, and objects mainly located towards the Galactic plane \citep{Rebassa2025}.

After performing more than one hundred realisations following the prescriptions described in Section~\ref{s:model}, we obtain a total of $13 \pm 5$ WD--BD systems within 100\,pc. This prediction is consistent with the currently known observed population, which contains 13 systems within the same distance limit. 
Although the comparison presented here is based on the currently known WD--BD population compiled from the literature rather than on a homogeneous $Gaia$-selected sample, we note that the inclusion of realistic $Gaia$ photometric selection cuts would substantially reduce the number of detectable synthetic systems. This illustrates the strong observational biases affecting the detection of WD--BD binaries, particularly at the faint end of the BD sequence. In fact, as stated in section \ref{s:obs}, only 3 of these 13 observed objects are detected as binaries by {\it Gaia}.

\subsection{The effects of the common envelope efficiency and the period distribution}
\label{ss:ce}

As discussed in the previous section, the total number of simulated WD--BD systems is very consistent with the observed population when adopting our standard set of model prescriptions. However, a more detailed analysis of the simulated binaries reveals that the predicted period distribution does not reproduce the observed sample satisfactorily, with close systems being significantly underrepresented in the simulations. 

Our reference simulation adopts a CE efficiency parameter of $\alpha_{\rm CE}=0.3$, following previous studies of WD binary populations (e.g. Zorotovic et al. 2010; Toonen \& Nelemans 2013; Santos-Garc\'{\i}a et al. 2025), which gives us a total of 1$\pm$1 close systems and 12$\pm$4 wide systems. Nevertheless, this value is too low, i.e. inefficient CE, to avoid having a large fraction of mergers and reproduce the observed number of short-period WD--BD systems, which are precisely the systems expected to have undergone a CE phase during their evolution. This may indicate that larger CE efficiencies are required in order to successfully reproduce close WD--BD binaries within population synthesis models. This result is broadly consistent with the CE efficiency studies of \citet{Ge+22,Ge+24} for short-period sdB+WD binaries. These works found that the CE ejection efficiency tends to increase with the initial mass ratio. Since WD--BD binaries are characterised by much larger values of $q=M_{\rm donor}/M_{\rm BD}$ than typical WD binaries, higher values of $\alpha_{\rm CE}$ may therefore be expected, facilitating their survival through the CE phase.

It is worth noting that the inferred CE efficiency also depends on the adopted treatment of the envelope binding energy. We recall that, in \texttt{MRBIN}, the envelope binding energy is calculated following the prescription of \citet{Claeys+14}, in which the effective envelope binding-energy parameter $\lambda$ is determined from the evolutionary properties of the donor star. Alternative treatments based on different stellar-evolution calculations can yield different envelope binding energies \citep[e.g.][]{Thai+26}, which may affect the survival probability of WD--BD systems during the CE phase and, consequently, the value of $\alpha_{\mathrm{CE}}$ required to reproduce the observed systems.

Regarding the initial orbital separation distribution, this can also introduce an additional uncertainty. A different separation distribution would modify the evolutionary stage at which systems enter the CE phase, and therefore the structure and binding energy of the donor envelope at CE onset, potentially leading to a different $\alpha_{\rm CE}$ required for WD--BD systems to survive. However, changes to the initial separation distribution would also affect the overall white dwarf binary population and must therefore remain consistent with the constraints of \citet{Torres+22}.

\begin{figure}[h!]
\centering
 \includegraphics[width=0.95\hsize]{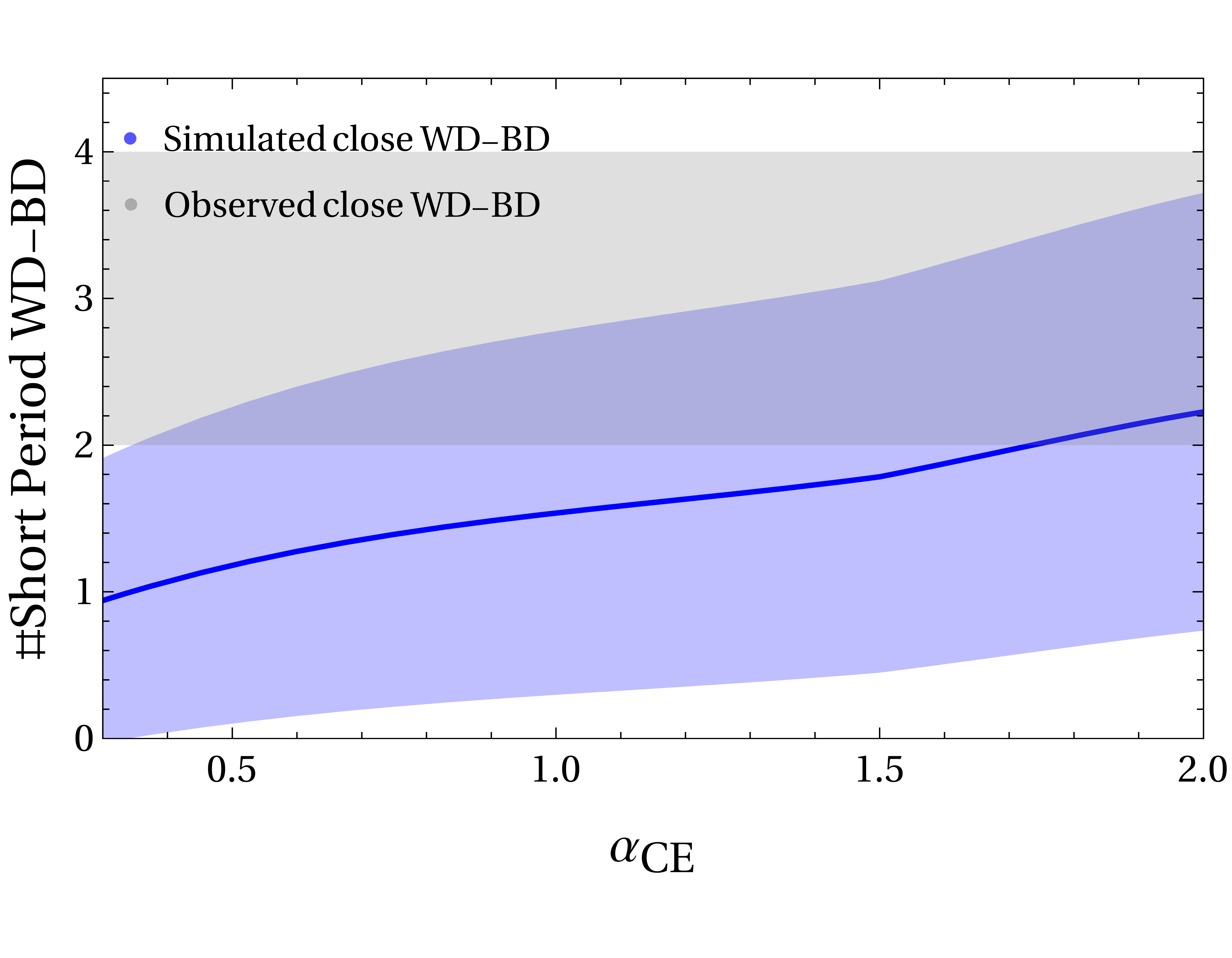}
 \caption{Number of simulated close WD--BD systems within 100\,pc as a function of the adopted common-envelope efficiency parameter $\alpha_{\mathrm{CE}}$. Increasing the CE efficiency significantly enhances the survival probability of short-period WD--BD binaries, improving the agreement with the observed population. 
 }
\label{FigCE}
\end{figure}

Figure~\ref{FigCE} illustrates how increasing the value of $\alpha_{\rm CE}$ leads to a corresponding increase in the number of close WD--BD systems produced by the simulations. Within the observed sample, 3 systems are identified as close PCEBs, including the system located at 101.6\,pc. Reproducing 3$\pm$1 close systems within the simulations requires $\alpha_{\rm CE} \gtrsim 0.4$ when considering the statistical uncertainties, while values above $\alpha_{\rm CE} \sim 1.8$ are needed to obtain a closer agreement with the observed sample. Values of $\alpha_{\rm CE}$ larger than 1 imply additional sources of energy apart from the orbital energy to eject the envelope.

The number of wide systems remains essentially unaffected by variations in $\alpha_{\rm CE}$, as expected given that these systems do not experience CE evolution. However, increasing the CE efficiency to reproduce the WD--BD population would also modify the predicted close populations of other binaries, such as WD-MS and double WD systems, which are already well reproduced using $\alpha_{\rm CE}=0.3$. This result suggests that adopting a single fixed value of the common envelope efficiency may not be appropriate when attempting to model multiple binary populations simultaneously.

\begin{figure}
\centering
\includegraphics[width=0.99\hsize]{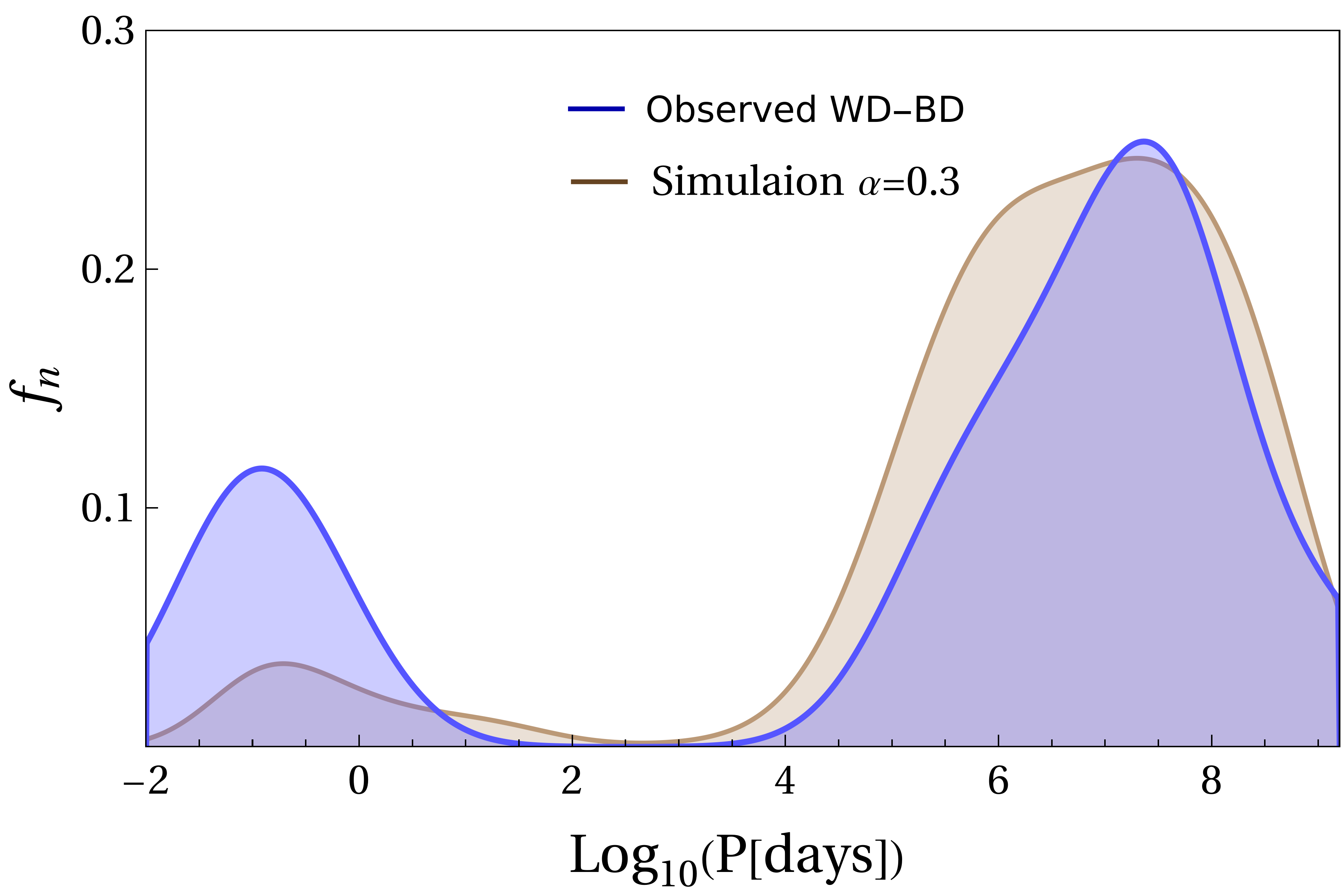}
\includegraphics[width=0.99\hsize]{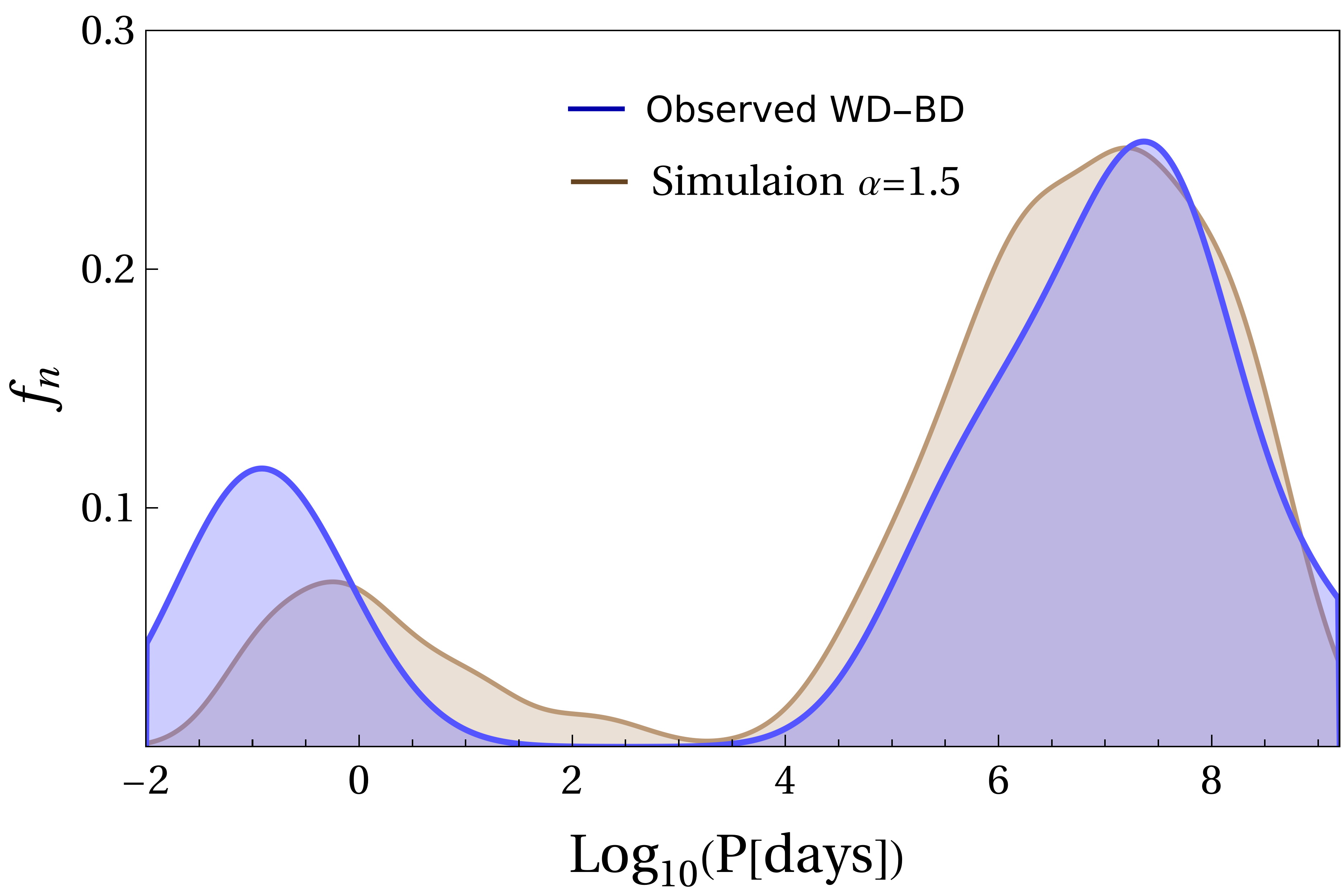}
\caption{Orbital period distribution for the simulated WD--BD population for two values of the CE efficiency. The simulations naturally reproduce a low-density region at intermediate orbital periods, consistent with the proposed WD--BD period gap.}
\label{F:per}
\end{figure}

We further investigate the orbital period properties of the observed WD–BD population and compare them with the distribution predicted by our synthetic sample. In Figure \ref{F:per}, we show the period distribution for the simulated population (brown) and the observed sample (blue) for an $\alpha_\mathrm{CE}$ value of 0.3 and 1.5. Both plots show that, as expected, the distribution of wide systems is not significantly affected by changes in the CE efficiency. Increasing $\alpha_{\rm CE}$, however, increases the number of short-period WD--BD systems, improving the agreement with the observed number of close systems. At the same time, higher values of $\alpha_{\rm CE}$ shift the predicted short-period systems towards longer orbital periods than those observed. In contrast, our reference value of $\alpha_{\rm CE}=0.3$ results in a distribution whose peak is better aligned with the observed one, although it underpredicts the number of short-period systems. In this context, a distribution of $\alpha_{\rm CE}$ values as in \citet{Torres+25}, or a prescription in which progenitors of WD--BD binaries preferentially evolve with larger CE efficiencies \citep{Zorotovic+22}, could provide a more realistic framework for population synthesis studies aiming to reproduce a broad range of binary populations rather than a single specific class of systems.

A noticeable feature in both the synthetic distribution and the observed sample is the presence of a reduced-density region at intermediate orbital periods, broadly consistent with the so-called “period gap” reported in previous studies \citep{Chen+26,  Nordhaus+13}. In our models, this gap arises naturally from binary evolution: systems undergoing CE evolution are driven to short orbital periods, while systems that avoid this phase experience orbital widening due to mass loss, populating the long-period regime. As a consequence, systems with intermediate orbital periods are intrinsically disfavoured. The observed WD–BD binaries are broadly consistent with this behaviour. While the intermediate-period region appears lightly populated, it is not completely devoid of systems, indicating that the gap should be interpreted as a region of low probability rather than a strict exclusion zone. In other words, the formation of systems in this range is possible, but significantly less likely within the current evolutionary framework and given the present-day sample size.

Overall, the data support the presence of a broad period gap rather than a sharp cutoff, fully consistent with the expectations from binary evolutionary theory.

We also quantify the overall agreement in period space by comparing the cumulative distribution functions (CDFs) of the observed and synthetic populations (Figure \ref{f:KS_Per}) using an $\alpha$ value of 1.5. Both distributions show very good consistency across the full range of orbital periods, with a Kolmogorov--Smirnov test yielding a p-value of 0.743, indicating no statistically significant difference between them. Some differences are nevertheless visible at short orbital periods, where the observed sample shows a slightly steeper rise in the CDF compared to the synthetic population, although this remains within statistical uncertainties given the small number of systems. At longer periods, both distributions converge towards unity, with no significant tension.

\subsection{Comparison of the observed and simulated WD--BD populations}
\label{ss:analysis}

To assess the agreement between the models and observations, we compare the observed WD--BD systems with the parameter distributions predicted by our simulations. These synthetic distributions were constructed from simulations computed with different values of the CE efficiency parameter, $\alpha_{\rm CE}$, ranging from 0.3 to 2. This approach allows us to include models capable of reproducing the observed population of close WD--BD systems, which are otherwise underrepresented when adopting the standard value of $\alpha_{\rm CE}=0.3$ alone.

\begin{figure}[h!]
\centering
\includegraphics[width=0.9\hsize]{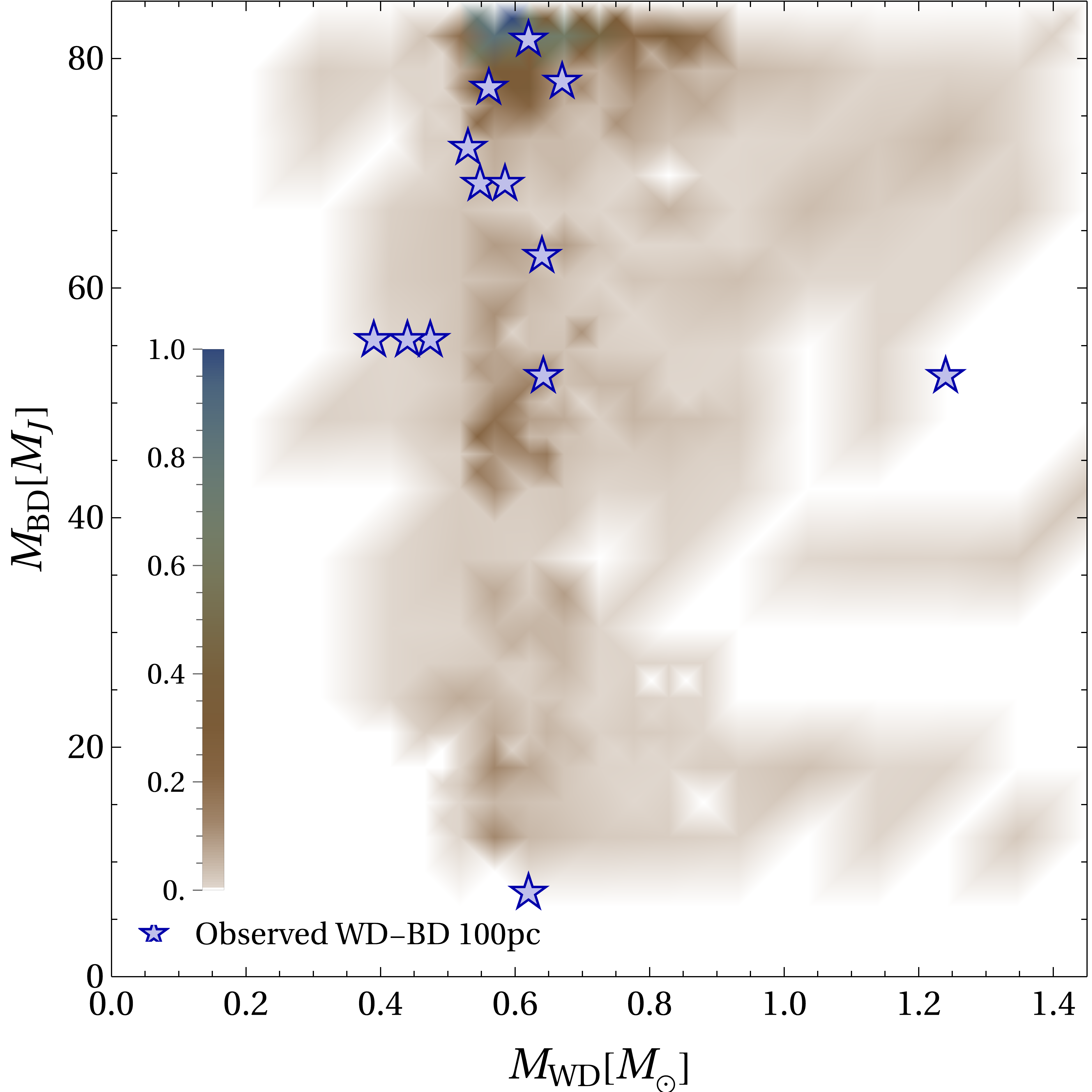}
\caption{Density distribution of the simulated WD--BD population in the white dwarf mass versus brown dwarf mass plane. Observed systems are shown as blue stars. Most observed systems lie within the highest-density regions predicted by the simulations, while a few outliers correspond to peculiar or extreme systems.}
\label{Figm}
\end{figure}

In Figure \ref{Figm}, we show the WD mass vs. BD mass plane. Most observed systems (represented as blue stars) lie within, or close to, the highest-density region of the distribution, while two of them fall outside the density map. One of these, WOLF 1130, hosts a WD with a mass of $1.2\,M_\odot$ and, as discussed in Section \ref{s:obs}, it belongs to a hierarchical system in which the WD has evolved as part of a binary with an M subdwarf, forming a triple system together with the BD companion WOLF 1130C. In this case, the WD–BD mass relation is not expected to be meaningful.

The other outlier corresponds to a system hosting an extremely low-mass BD with $M = 0.007\,M_\odot$, below our adopted lower mass limit of $0.01\,M_\odot$. In addition, there are three systems containing low-mass He-core WDs ($M_{\rm WD} < 0.5\,M_\odot$), whose BD companions all have masses around $0.053\,M_\odot$ and lie within the highest-density region of the distribution for $M_{\rm WD} < 0.5\,M_\odot$.

\begin{figure}
\centering
\includegraphics[width=0.99\hsize]{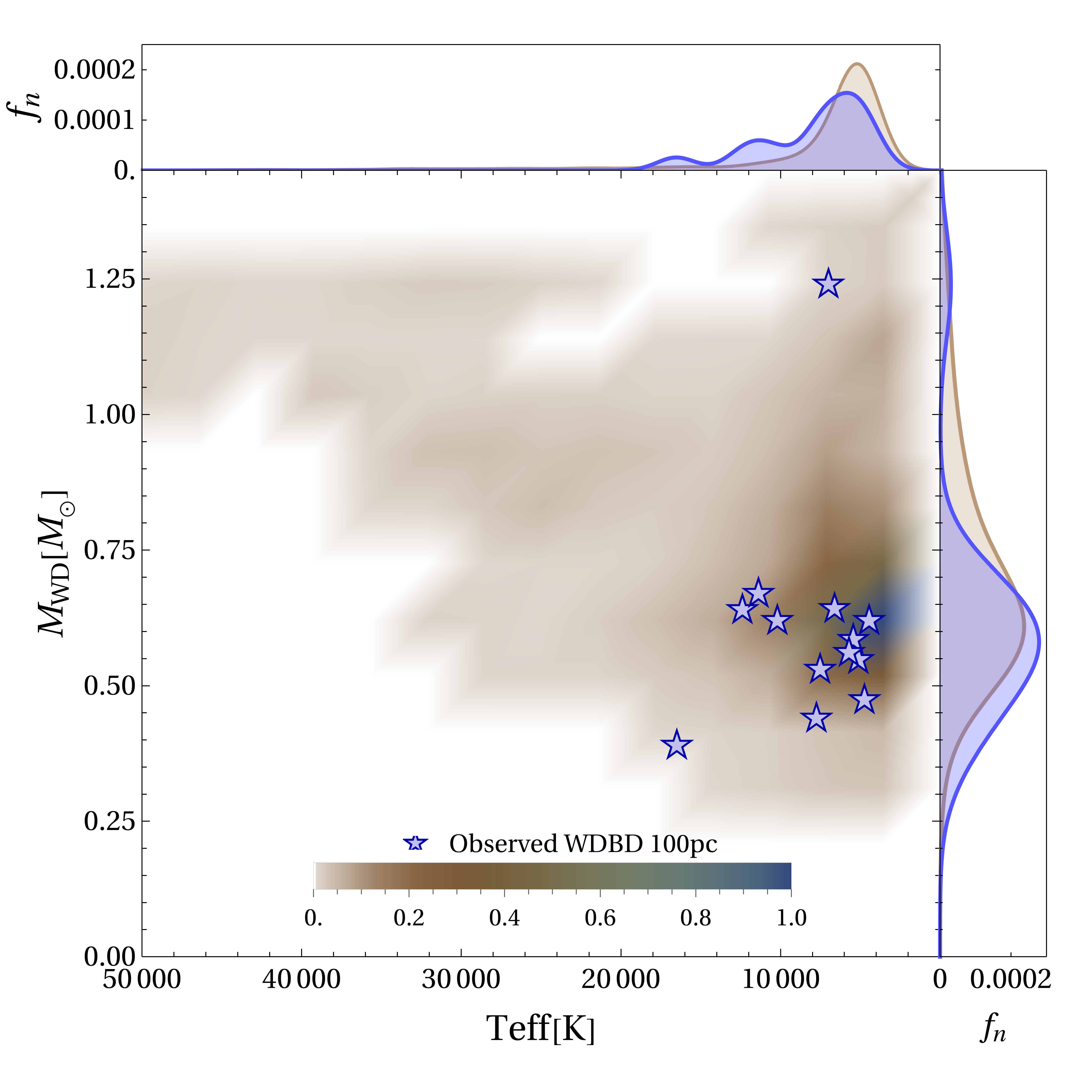}
\includegraphics[width=0.99\hsize]{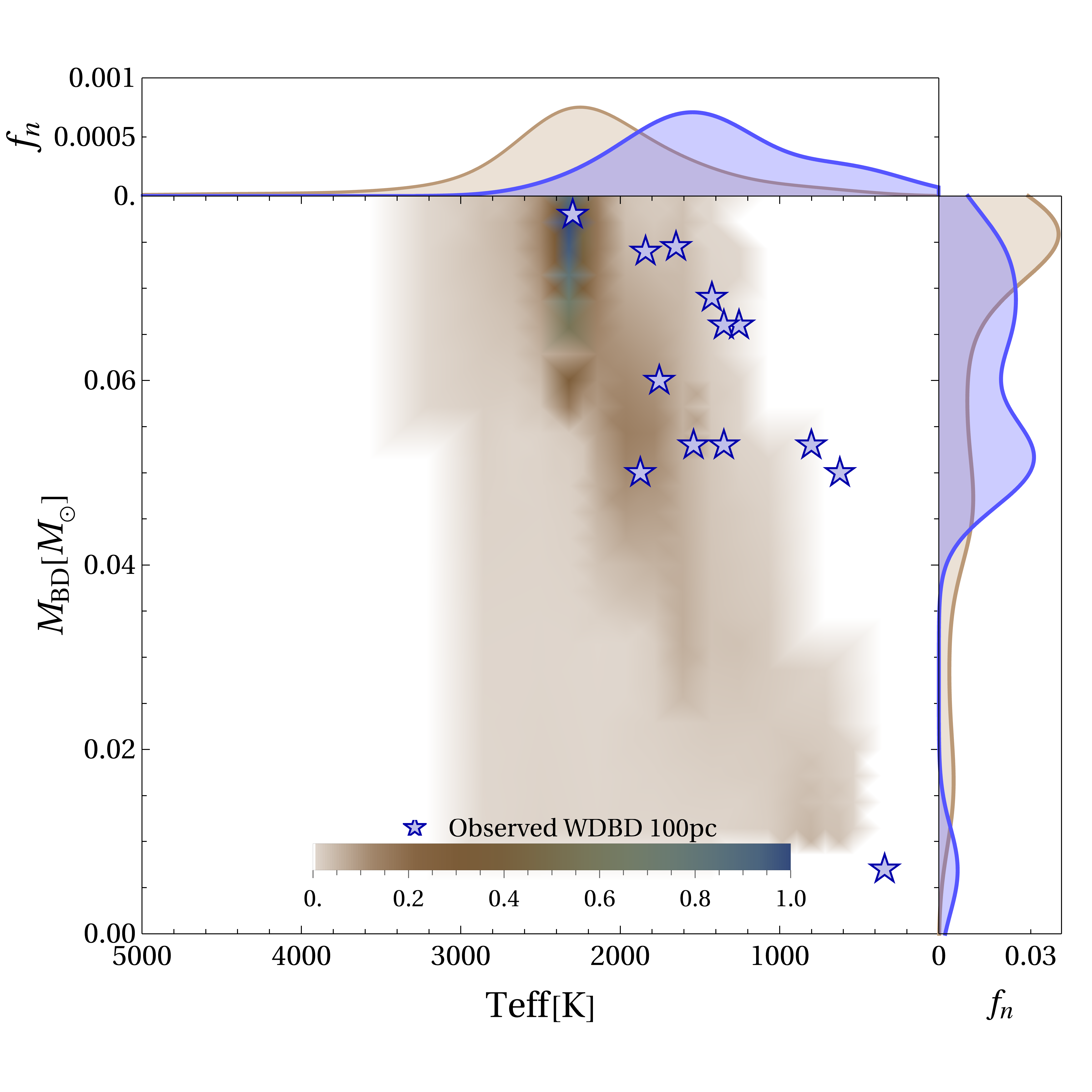}
\caption{Mass-effective temperature distributions of the simulated and observed WD--BD populations. Top panels: white dwarf component. Bottom panels: brown dwarf component. The synthetic populations are represented as density maps, while the observed systems are shown as blue stars. The side and upper panels display the corresponding one-dimensional distributions for the simulated (brown) and observed (blue) samples.}
\label{Figmteff}
\end{figure}

Figure~\ref{Figmteff} shows the mass-effective temperature distributions for both the WD (top panel) and BD (bottom panel) components of the simulated and observed WD--BD populations. The synthetic populations are represented as density maps, while the observed systems are indicated by blue stars. In each case, the corresponding one-dimensional distributions of mass and effective temperature are also displayed in the side and upper panels for the simulated (brown) and observed (blue) samples. The synthetic distributions combine the results from 100 independent realisations, allowing for a more robust statistical characterisation of the simulated WD--BD population, while the observed distributions are shown using kernel density estimates to mitigate small-number fluctuations.

The WD distributions show an apparent good agreement between the simulations and the observed sample, although the observed mass distribution appears slightly shifted towards lower masses. Most of the observed WDs are located within the highest density regions of the synthetic distribution, indicating that the models successfully reproduce the most probable parameter space occupied by these systems. In addition, one low-mass, high-temperature WD is also found within the simulated density map, although in a region associated with a significantly lower probability. Reproducing this low-mass, high-temperature region requires adopting larger values of $\alpha_{\rm CE}$, as higher CE efficiencies increase the survival probability of close systems containing low-mass WDs. The two samples are further analysed via comparing the corresponding cumulative distributions, which can be insepected in  Figure \ref{f:KS_MTeff_WD}. 

To quantify the similarity between the observed and simulated distributions, we performed a Kolmogorov--Smirnov (KS) test on the cumulative WD mass distribution. The resulting p-value, $p=0.143$, is larger than the commonly adopted significance threshold of 0.05. Therefore, the null hypothesis that both samples are drawn from the same underlying distribution cannot be rejected. Although the agreement is not statistically strong, the test indicates that the simulated and observed WD mass distributions remain compatible within the limitations imposed by the small observational sample.

A similar analysis was performed for the WD effective temperatures, whose cumulative distributions are shown in the right panel of Figure~\ref{f:KS_MTeff_WD}. In this case, the KS test yields a p-value of $p=0.088$. While still above the standard significance threshold, this lower value suggests a slightly weaker agreement between the observed and simulated temperature distributions. Nevertheless, the result remains statistically consistent with the hypothesis that both samples originate from the same parent population. 

An analogous analysis was performed for the BD component, shown in the bottom panel of Figure~\ref{Figmteff}. In this case, the effective temperatures of the observed BDs tend to be systematically lower than those predicted by the simulations, with some observed systems lying outside the highest-density regions of the synthetic distribution. This discrepancy is more clearly illustrated in the right panel of Figure~\ref{f:KS_MTeff_BD}, which presents the CDFs of the BD effective temperatures. Applying a KS test to both samples yields a very low p-value of $2.9 \times 10^{-6}$, indicating that the observed and simulated temperature distributions are statistically inconsistent and are unlikely to originate from the same parent population. The origin of this discrepancy remains uncertain, although it may partly reflect the considerable observational challenges involved in determining the physical properties of BDs, particularly their effective temperatures and masses. The temperatures assigned to the simulated BDs are also model dependent, as they are determined from evolutionary models for a given mass and age. Different evolutionary grids can predict different temperatures, which introduces an additional systematic uncertainty when comparing the simulated and observed BD temperature distributions. As a simple numerical experiment, we find that introducing an ad hoc systematic offset of approximately 550--700\,K in the simulated temperatures leads to a significantly better agreement between both populations and substantially improves the resulting p-values, as illustrated in Figure~\ref{f:KS_Teff+700}.

The BD mass distributions show a somewhat better agreement. However, the simulated population exhibits a clear peak around $\sim0.075\,M_{\odot}$, whereas the observed systems appear more uniformly distributed between $\sim0.05$ and $0.08\,M_{\odot}$. The corresponding CDFs and KS test results are shown in the left panel of Figure~\ref{f:KS_MTeff_BD}. In this case, the KS test yields a p-value of $p=0.049$. Although this value is still above the standard significance threshold of 0.05, suggesting that the null hypothesis cannot be rejected, the relatively low p-value points towards a possible tension between the observed and simulated BD mass distributions. As for the effective temperatures, part of this discrepancy may be related to the substantial observational uncertainties affecting the determination of BD masses.

\section{The simulated MS--BD population and the BD desert}
\label{ss:bddesert}

Although the present work does not explicitly assume the existence of the so-called BD desert, our synthetic population naturally includes main-sequence–brown dwarf (MS–BD) binaries, allowing us to test whether this feature emerges self-consistently from the formation and evolutionary channels implemented in the model. To compare with observations, we adopt the most recent and most complete MS–BD catalogue, compiled by \citet{Chen+26}, which currently represents the largest homogeneous sample of such systems with a total of 159 with measured masses and periods. The observational catalogue is not volume limited and contains systems spanning a wide range of distances, whereas our simulations correspond to a volume-limited sample within 100\,pc. Therefore, the comparison should be regarded as qualitative rather than strictly statistical.
This catalogue already suggests a more nuanced picture of the BD desert than the traditionally assumed near-absence of companions in the $\sim 0.04,M_\odot$ regime \citep{Grether+06,MaGe14,Stevenson+23} while the number of objects is indeed reduced around this mass scale, systems are still present, indicating that the “desert” is at most a depletion rather than a complete lack of companions.

\begin{figure}
\centering
\includegraphics[trim=0 0 0 0 clip=true,width=0.95\hsize]{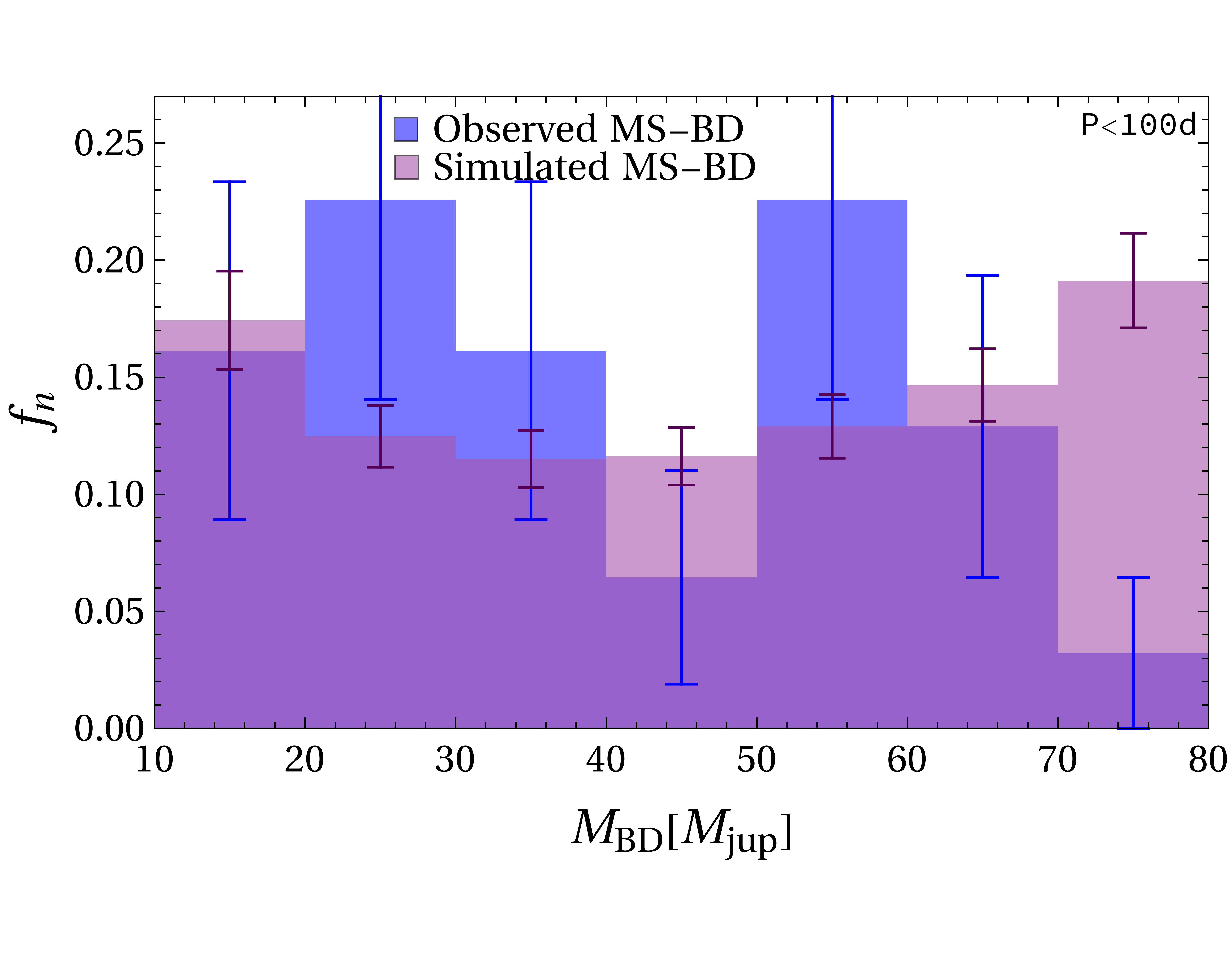}
\includegraphics[trim=0 0 0 0 clip=true, width=0.95\hsize]{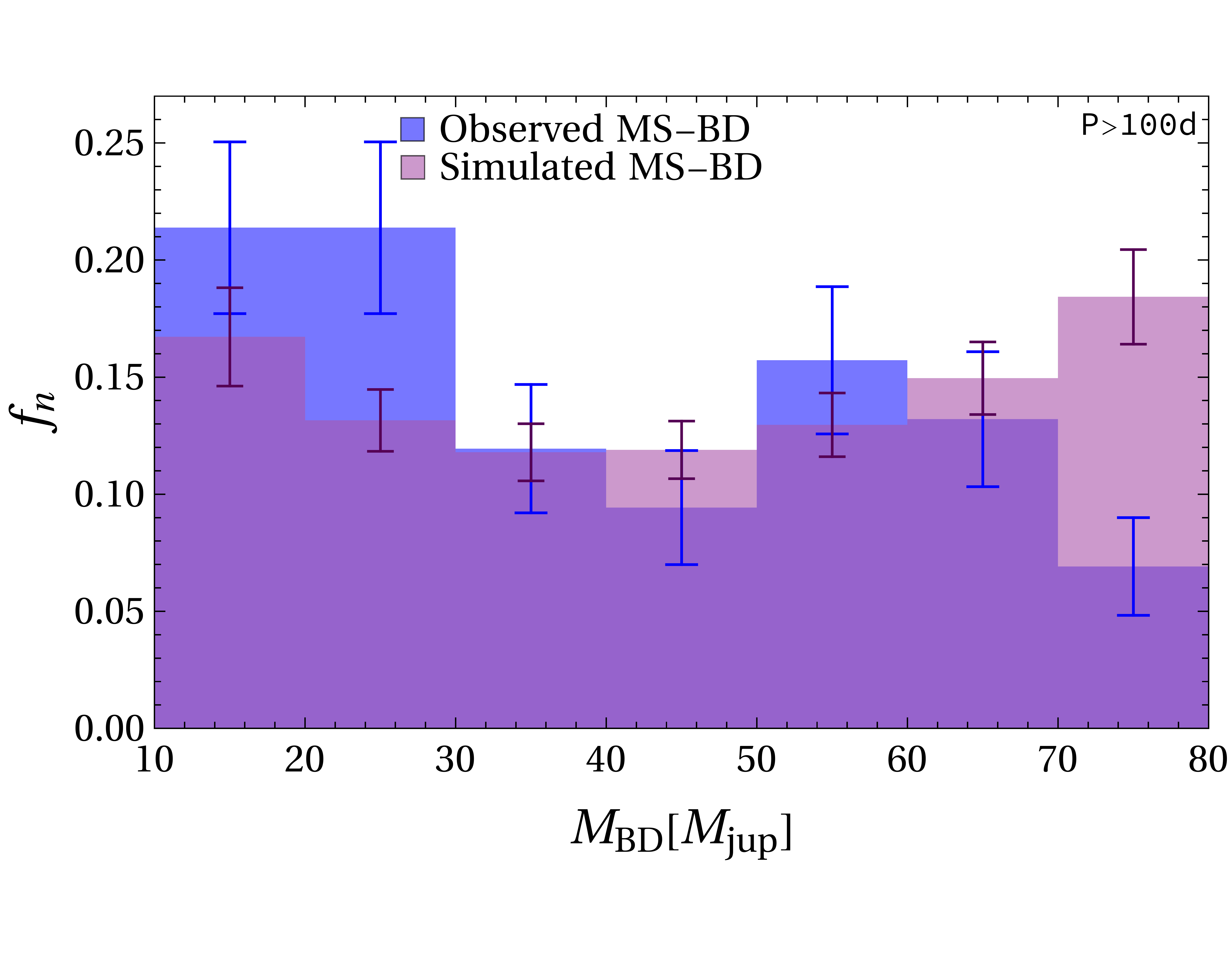}
\caption{Mass distribution of main-sequence--brown dwarf binaries from the simulations and the observational sample of \citet{Chen+26} Top panel: short-period systems ($P < 100$ days). Bottom panel: long-period systems ($P > 100$ days). Both the observations and simulations show a relative depletion around $\sim0.04\,M_{\odot}$, consistent with a weakened brown dwarf desert.}
\label{f:msbd}
\end{figure}

In Figure \ref{f:msbd} we compare the observed BD mass distribution of MS–BD systems (blue) with our simulated population (magenta), separating short-period systems ($P < 100$ days, top panel) from longer-period systems ($P > 100$ days, bottom panel) since previous studies \citep{Stevenson+23} suggest that the BD desert appears in the mass distribution of systems with periods smaller than 100 days. The simulated sample shows a broad distribution extending across the full BD mass range ($\sim 10$–$80\,M_{\mathrm{Jup}}$), with a mild underdensity around $\sim 30$–$50\,M_{\mathrm{Jup}}$ for both ranges of periods.

The observed sample for periods smaller than 100\,days shows a clear underpopulation between $40$--$50\,M_{\mathrm{Jup}}$ and for masses higher than $70\,M_{\mathrm{Jup}}$, with two peaks at $20$--$30\,M_{\mathrm{Jup}}$ and $50$--$60\,M_{\mathrm{Jup}}$. Our simulations reproduce several qualitative features of the observed distribution. In particular, the simulated BD mass function is not uniform, displaying a relative enhancement towards both the low-mass end ($\lesssim 20\,M_{\mathrm{Jup}}$) and the high-mass end ($\gtrsim 60\,M_{\mathrm{Jup}}$). Between these two regimes, the distribution exhibits a broad relative minimum around $M_{\mathrm{BD}}\sim0.04\,M_\odot$ ($\sim40\,M_{\mathrm{Jup}}$), which may simply reflect the shape of the underlying mass distribution rather than a distinct physical feature.

We also note that the observed binary fraction decreases towards the stellar/substellar boundary, whereas the simulated population shows a tendency towards higher binary fractions at larger brown dwarf masses. Given the relatively large statistical uncertainties, however, the simulations do not provide strong evidence for a significant increase in the binary fraction. This difference may reflect uncertainties in the adopted initial binary parameters, particularly the mass-ratio distribution, or in the evolutionary treatment of low-mass companions.

To quantify the level of agreement between the observed and simulated BD mass distributions, we performed two KS tests separately for short- and long-period systems. For systems with $P < 100$\,d, we obtain a p-value of $ 9.8 \times 10^{-2}$, indicating that the null hypothesis that both samples are drawn from the same underlying distribution cannot be rejected at the 5\% significance level. In contrast, for $P > 100$\,d we find a p-value of  $0.001$, providing strong evidence that the simulated and observed distributions differ significantly in this regime.

Overall, both the observational catalogue and our simulations support a scenario in which the BD desert is not a true absence of objects, but rather a relative depletion whose depth and sharpness depend on orbital period and formation pathway. In this context, the agreement between observations and simulations is encouraging, as it suggests that the implemented formation channels are capable of naturally producing a weakened but non-zero desert-like feature around 30-50 M$_{\rm Jup}$, without requiring it to be imposed ad hoc.

While the BD desert is primarily defined in terms of the companion mass and orbital period distribution, the systems that populate this region also carry information about their formation and dynamical history. In this context, additional orbital parameters such as eccentricity provide complementary constraints on the physical processes that shaped the surviving MS–BD binaries.

\begin{figure}
\centering
\includegraphics[trim=0 0 0 0 clip=true,width=0.95\hsize]{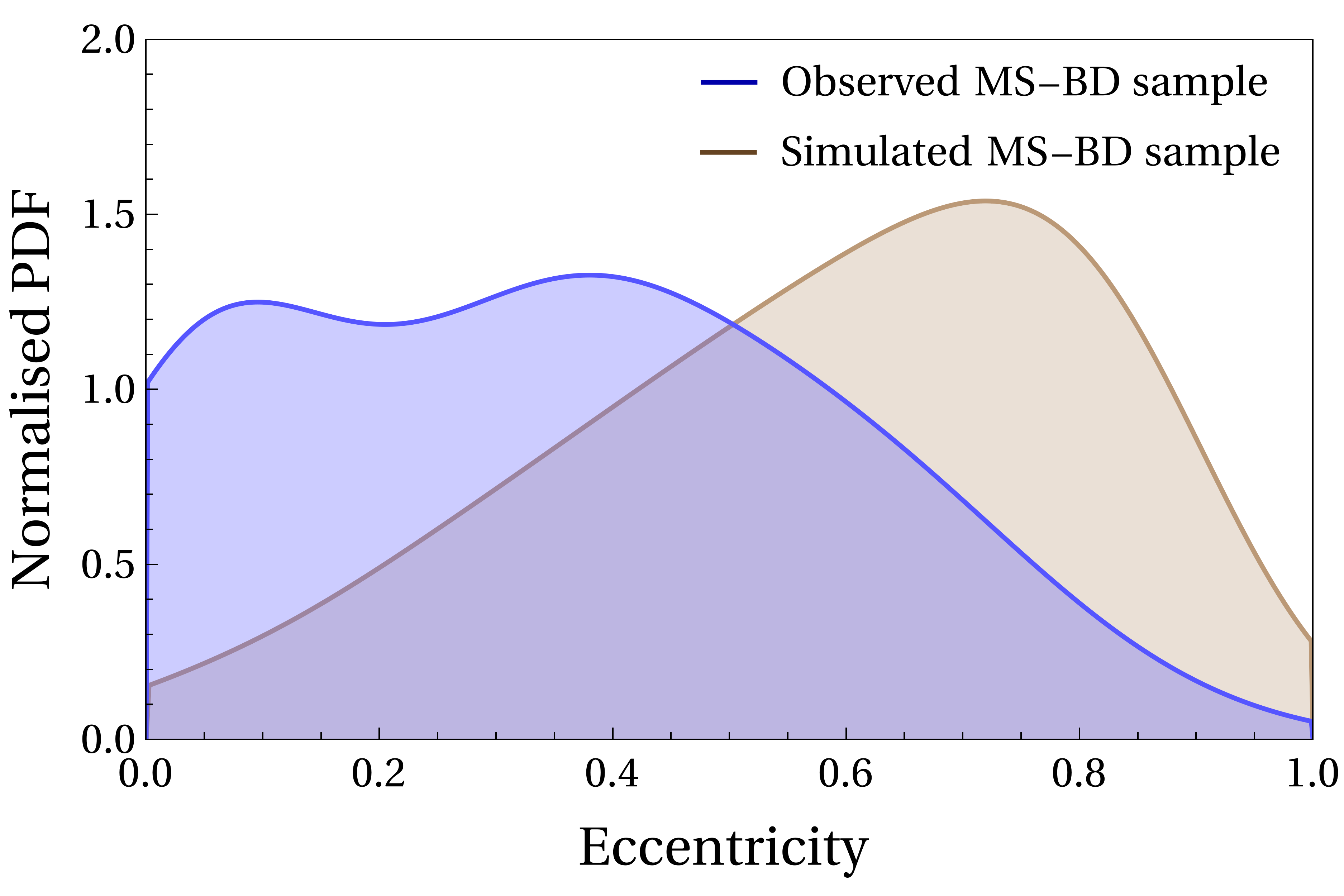}
\caption{Eccentricity distribution of MS--BD binaries predicted by our \texttt{MRBIN} simulations, compared with the observational sample of \citet{Chen+26}.}
\label{f:ecc}
\end{figure}

Following the approach of \citet{Bowler+20}, we compared the eccentricity distribution of MS–BD binaries in our simulations with the observational sample compiled by \citet{Chen+26}, as shown in Figure\,\ref{f:ecc}. The simulated population shows good qualitative agreement with the eccentricity distribution reported in \citet{Bowler+20}, including a similar shape and a preference for high-eccentricity systems.
In contrast, the comparison with the more recent and homogeneous MS–BD sample of \citet{Chen+26} reveals a different behaviour, with a flatter distribution at low to intermediate eccentricities ($0 \lesssim e \lesssim 0.4$) and a less pronounced high-eccentricity tail than predicted by the simulations.
These results suggest that, while the simulations reproduce the general orbital architecture of previously studied samples, differences remain when compared to more homogeneous and recent observational datasets, highlighting ongoing uncertainties in the formation and evolution of low-mass binary systems.

\section{Conclusions}

In this work we have extended the \texttt{MRBIN} binary population synthesis code into the substellar regime by incorporating evolutionary models for brown dwarfs (BDs) and adopting an updated initial mass function at very low masses. This allowed us to consistently model populations containing single BDs, BD--BD binaries, and white dwarf--brown dwarf (WD--BD) systems within the solar neighbourhood.

Using this updated framework, we simulated the WD--BD population within 100\,pc of the Sun and compared the results with the currently known observational sample. The simulations predict a total of $13 \pm 5$ WD--BD systems within this volume, in remarkable agreement with the $13 \pm 4$ systems currently known from observations. This implies that WD--BD binaries represent only $\sim0.1\%$ of the local white dwarf population, highlighting their observed scarcity. Importantly, this result is obtained using the same Galactic and binary population prescriptions previously shown to reproduce the observed WD population and several classes of WD binaries \citep{Torres+22}, without any additional tuning aimed specifically at the WD--BD population. The ability of the model to simultaneously reproduce both the local WD population and the observed number of WD--BD systems suggests that the apparent scarcity of these binaries can be naturally explained within current binary population synthesis frameworks once substellar companions are consistently included.

The $Gaia$ Hertzsprung--Russell diagram derived from our simulations reveal a relatively well-defined region where unresolved WD--BD binaries are expected to lie. However, this region is strongly contaminated by sources with poor photometric or astrometric quality, making observational identification difficult. We also show that $Gaia$ searches for these systems are significantly more reliable in the $G-RP$ colour space than in BP-RP, where the BP photometry becomes highly uncertain at faint magnitudes.

A detailed comparison between the observed and simulated parameter distributions shows generally good agreement for the WD component. The distributions of WD masses, effective temperatures, and orbital periods are statistically consistent with the observations according to KS tests. In contrast, the BD effective temperatures predicted by the models appear systematically hotter than those inferred observationally. A temperature shift of approximately 550--700\,K substantially improves the agreement, suggesting possible systematic uncertainties either in the adopted substellar evolutionary models or in the observational estimates of BD atmospheric parameters.

Our simulations reproduce a broad low-density region at intermediate orbital periods, consistent with the proposed WD--BD period gap. This feature emerges naturally from binary evolution: systems that undergo common-envelope (CE) evolution evolve towards short orbital periods, while systems avoiding interaction migrate towards wider separations as a consequence of mass loss. As a result, the intermediate-period region is not completely devoid of systems, but is intrinsically less populated. The detailed distribution of close binaries, however, depends sensitively on the adopted CE efficiency. Using the standard value $\alpha_{\rm CE}=0.3$, commonly adopted in studies of WD binaries, the simulations significantly underproduce the observed population of short-period WD--BD systems. Reproducing these close binaries requires larger efficiencies, with $\alpha_{\rm CE}\gtrsim0.5$ and potentially values approaching $\alpha_{\rm CE}\sim1.8$. Such large efficiencies, however, would modify the predicted populations of other close binaries that are already well reproduced with lower values of $\alpha_{\rm CE}$. This result supports the idea that a single universal CE efficiency may not be sufficient to simultaneously reproduce all binary systems \citep{Torres+25}, and suggests that WD--BD progenitors may preferentially evolve under different CE conditions or through a broader distribution of CE efficiencies.

The relatively high CE efficiencies required in our simulations should be interpreted within the adopted population-synthesis framework. Alternative prescriptions for the envelope binding energy or the initial semi-major-axis distribution could increase the survival probability of WD--BD systems through the CE phase and thus reduce the required $\alpha_{\rm CE}$. However, these prescriptions are also constrained by the agreement of the model with other observed WD binary populations \citep{Torres+22,Santos+25}, and changing them would require a re-calibration of the overall binary population.

Finally, our simulations of main-sequence--brown dwarf binaries naturally reproduce a weakened version of the BD desert without explicitly imposing it in the models. Both the observed and synthetic populations show a relative depletion of companions around $\sim 0.04\,M_{\odot}$ rather than a complete absence of systems, supporting the interpretation of the BD desert as a statistical underdensity linked to formation and evolutionary processes.

The predicted mass distribution of the BD companions is sensitive to the adopted initial mass-ratio distribution. We adopt $n(q)\propto q^{-1.13}$ following \citet{Torres+22}, who found this prescription to provide the best agreement with the observed white dwarf binary population when simultaneously constraining the initial binary parameters. However, the low-$q$ regime of the mass-ratio distribution is particularly relevant for the formation of both MS--BD and WD--BD systems. \citet{MoeDiStefano17} found evidence for a change in the mass-ratio distribution below $q\simeq0.3$, with the distribution becoming increasingly weighted towards higher mass ratios. Although the Moe \& Di Stefano prescription was also considered in \citet{Torres+22}, the $q^{-1.13}$ model provided the best overall agreement with the observational constraints considered in that study. Adopting the Moe \& Di Stefano prescription would modify the initial binary population and, consequently, the calibration of the model against the other observed WD binary populations. A meaningful comparison would therefore require a re-calibration of the population-synthesis model rather than an isolated change to the mass-ratio distribution. Nevertheless, the uncertain behaviour of the mass-ratio distribution in the low-$q$ regime remains an important source of uncertainty in our predictions of the WD--BD population.

Overall, our results demonstrate the importance of extending binary population synthesis studies into the substellar regime. The ability of the updated \texttt{MRBIN} population synthesis code to simultaneously reproduce both the observed WD population and the currently known WD--BD systems demonstrates that the apparent scarcity of WD--BD binaries can be naturally explained within standard binary evolution scenarios once substellar companions are consistently included. At the same time, the properties of the observed close WD--BD population provide valuable constraints on common-envelope evolution in the extreme mass-ratio regime. Future observational surveys, particularly those combining \textit{Gaia} astrometry with infrared facilities, will be crucial for improving the census of WD--BD systems and further constraining the formation and evolution of compact binaries containing substellar companions.

\begin{acknowledgements}

This work was partially supported by the Spanish MINECO grant PID2023-148661NB-I00 and by the AGAUR/Generalitat de Catalunya grant SGR-386/2021, and by the MINECO grant PID2020-117252GB-I00 and the PhD grant PRE2021-100503 funded by MICIU/AEI/10.13039/501100011033 and ESF+. HG acknowledges the National Natural Science Foundation of China (No. 12525304), the National Key R\&D Program of China (No. 2021YFA1600403), the Strategic Priority Research Program of the Chinese Academy of Sciences (No. XDB1160201), and the CAS Project for Young Scientists in Basic Research (YSBR-148).
This work has made use of data from the European Space Agency (ESA) mission {\it Gaia} (\url{https://www.cosmos.esa.int/gaia}), processed by the {\it Gaia} Data Processing and Analysis Consortium (DPAC, \url{https://www.cosmos.esa.int/web/gaia/dpac/consortium}). Funding for the DPAC has been provided by national institutions, in particular the institutions participating in the {\it Gaia} Multilateral Agreement.
\end{acknowledgements}
\section*{Data Availability Statement}
The data underlying this article are available in the article.  Supplementary material will be shared on reasonable request to the corresponding author.


\bibliographystyle{aa}
\bibliography{WDBD}

\appendix
\onecolumn

\section{Observed WD--BD sample in \emph{Gaia}}
\label{a:observedgaia}

In table \ref{Tab2} we present the \textit{Gaia} magnitudes found for the components of the WD--BD binary systems of study for this work. All of the magnitudes have been obtained by checking their coordinates in the \textit{Gaia} DR3 catalogue. Only one of the BD components have been found by searching for proper motion pairs close to each of the WDs. However, apparent magnitudes for some of the BD have been found in different catalogues, giving us an idea of how faint these kind of objects can be.

\begin{table*}[h!]
\centering
\caption{{\it Gaia} magnitudes for known WD--BD binary systems used for comparison with our population synthesis models.}
\label{Tab2}
\begin{tabular}{c|lllll}
\textbf{Object}                  & \textbf{Component}        & \textbf{{\it Gaia} ID }             & \textbf{$G$ (mag)}          & \textbf{$G_{\rm BP}$ (mag) }    & \textbf{$G_{\rm RP}$ (mag) }    \\ \hline     \hline
\multirow{2}{*}{WOLF 1130}            & BD          & Not found            & -          & -      & -      \\
                                      & WD          & 2185710338703934976  & 11.17      & 12.23  & 10.14  \\    \hline
\multirow{2}{*}{WD 0806-661}          & BD          & Not found            & 20 (opt)   & -      & -      \\
                                      & WD          & 5274517467840296832  & 13.665     & 13.682 & 13.616 \\    \hline   
\multirow{2}{*}{LSPM J0055+5948}      & BD          & Not found            & -          & -      & -      \\
                                      & WD          & 426122397136335872   & 16.806     & 17.362 & 16.117 \\    \hline
\multirow{2}{*}{COCONUTS-1}           & BD          & Not found            & -          & -      & -      \\
                                      & WD          & 244214799689691904   & 17.159     & 17.62  & 16.554 \\    \hline
\multirow{2}{*}{GD 165AB}             & BD          & Not found            & 20.8 (opt) & -      & -      \\
                                      & WD          & 1176717792385803136  & 14.344     & 14.34  & 14.369 \\    \hline
\multirow{2}{*}{LSPM J0806+2215}      & BD          & Not found            & 25 (SDSS)  & -      & -      \\
                                      & WD          & 677060225090183552   & 17.789     & 18.124 & 17.292 \\    \hline
PHL 5038AB                            & Double star & 2677851743291189888  & 17.327     & 17.483 & 17.096 \\    \hline
\multirow{2}{*}{LSPM J1459+0857}      & BD          & Not found            & 22 (opt)   & -      & -      \\
                                      & WD          & 1161820298887863424  & 19.001     & 19.454 & 18.497 \\    \hline
\multirow{2}{*}{VVV J1256-62AB}       & BD          & 5863122429178232704  & 20.7       & 21.822 & 19.24  \\
                                      & WD          & 5863122429179888000  & 19.58      & 20.22  & 18.899 \\    \hline
\multirow{2}{*}{LSPM J0241+2553}      & BD          & Not found            & -          & -      & -      \\
                                      & WD          & 126285095203324672   & 18.138     & 18.388 & 17.759 \\     \hline
\multirow{2}{*}{GD 1400}              & BD          & Not found            & -          & -      & -      \\
                                      & WD          & 5135466183642594304  & 15.211     & 15.232 & 15.178  \\    \hline
NLTT5306                              & Double star & 2588874825669925504  & 16.901     & 17.052 & 16.637 \\    \hline
\multirow{2}{*}{WD 0137-349}          & BD          & Not found            & -          & -      & -      \\
                                      & WD          & 5012047927570807552  & 15.346     & 15.284 & 15.485 \\    \hline
\end{tabular}
\
\end{table*}

\clearpage

\section{Kolmogorov--Smirnov tests}
\label{a:KStests}

\begin{figure*}[h!]
\centering
\includegraphics[trim=0 0 0 0 clip=true, width=0.49\columnwidth]{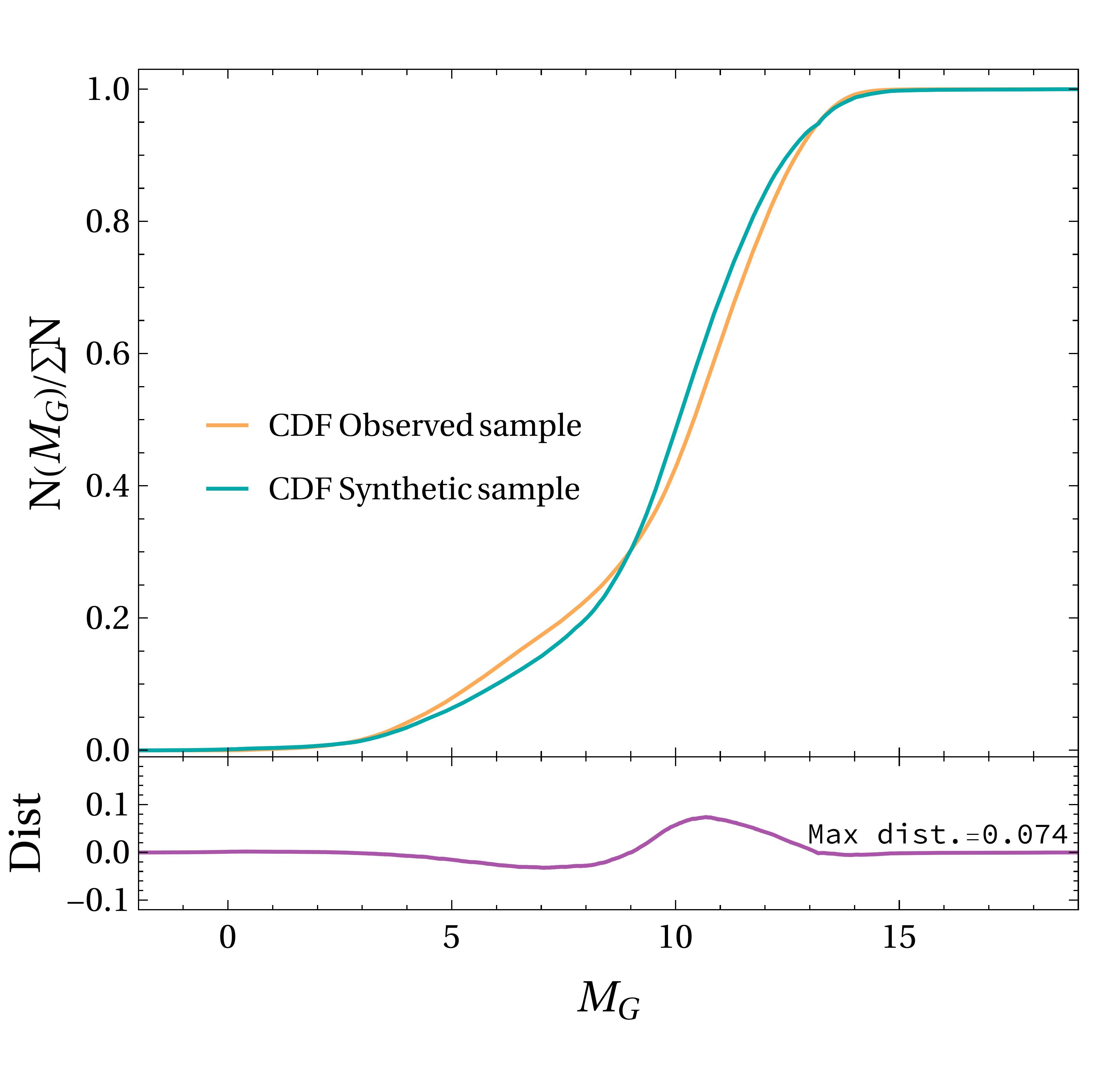}
\includegraphics[trim=10 8 10 0 clip=true, width=0.485\columnwidth]{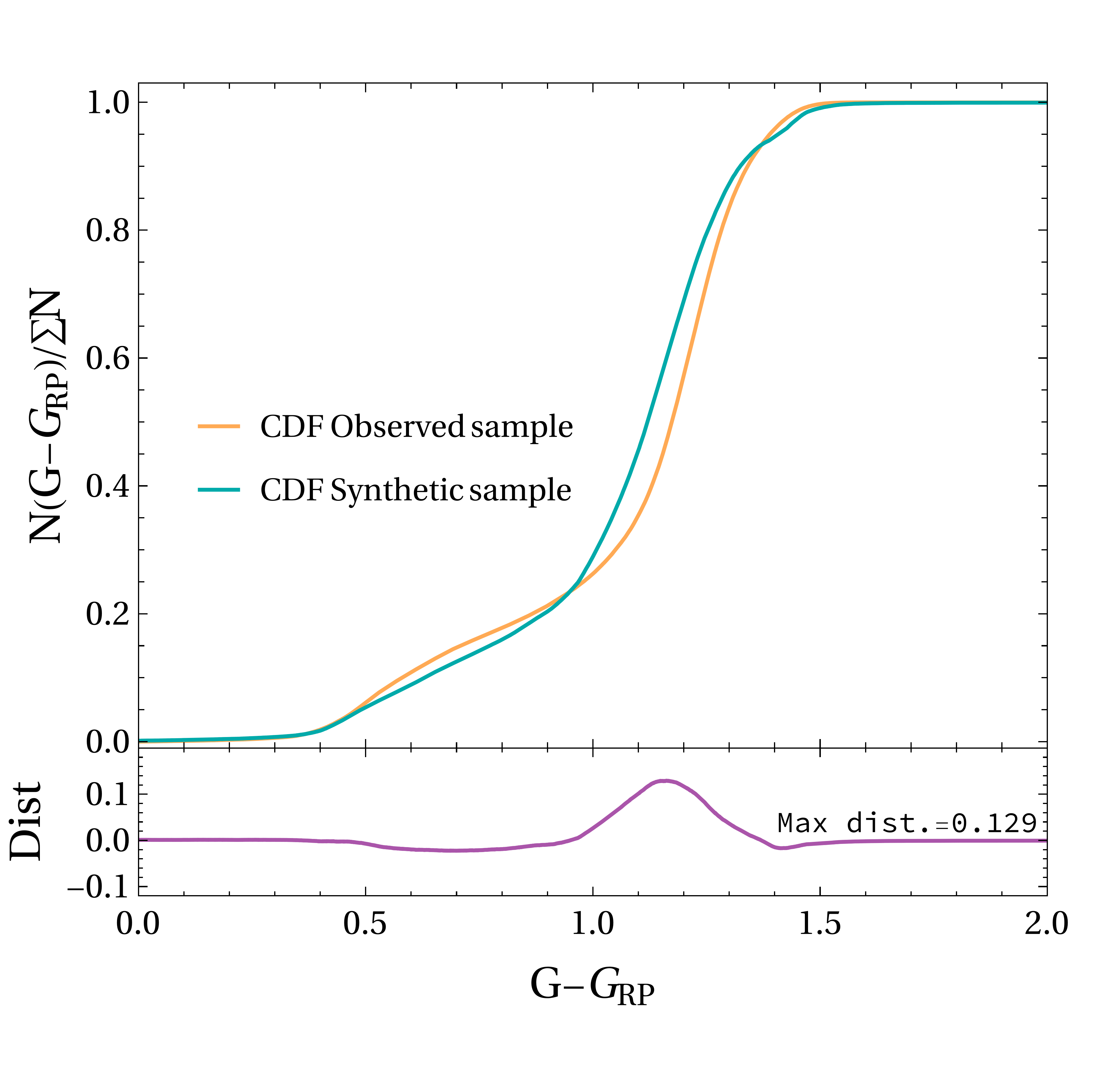}
\caption{ Cumulative Distribution Function (CDF) of the synthetic sample (cyan line) and the observed sample at 100\,pc (orange line) for the magnitude $M_G$ (left panel) and the color $G-G_{\rm RP}$ (right panel) after including BD models in our simulations. For a more quantitative comparison, the distance (DIST; magenta line) between the CDFs is also plotted.}
\label{f:KS_gaiaVSsim}
\end{figure*}

\begin{figure*}[h!]
\centering
\includegraphics[trim=10 8 10 0 clip=true, width=0.485\columnwidth]{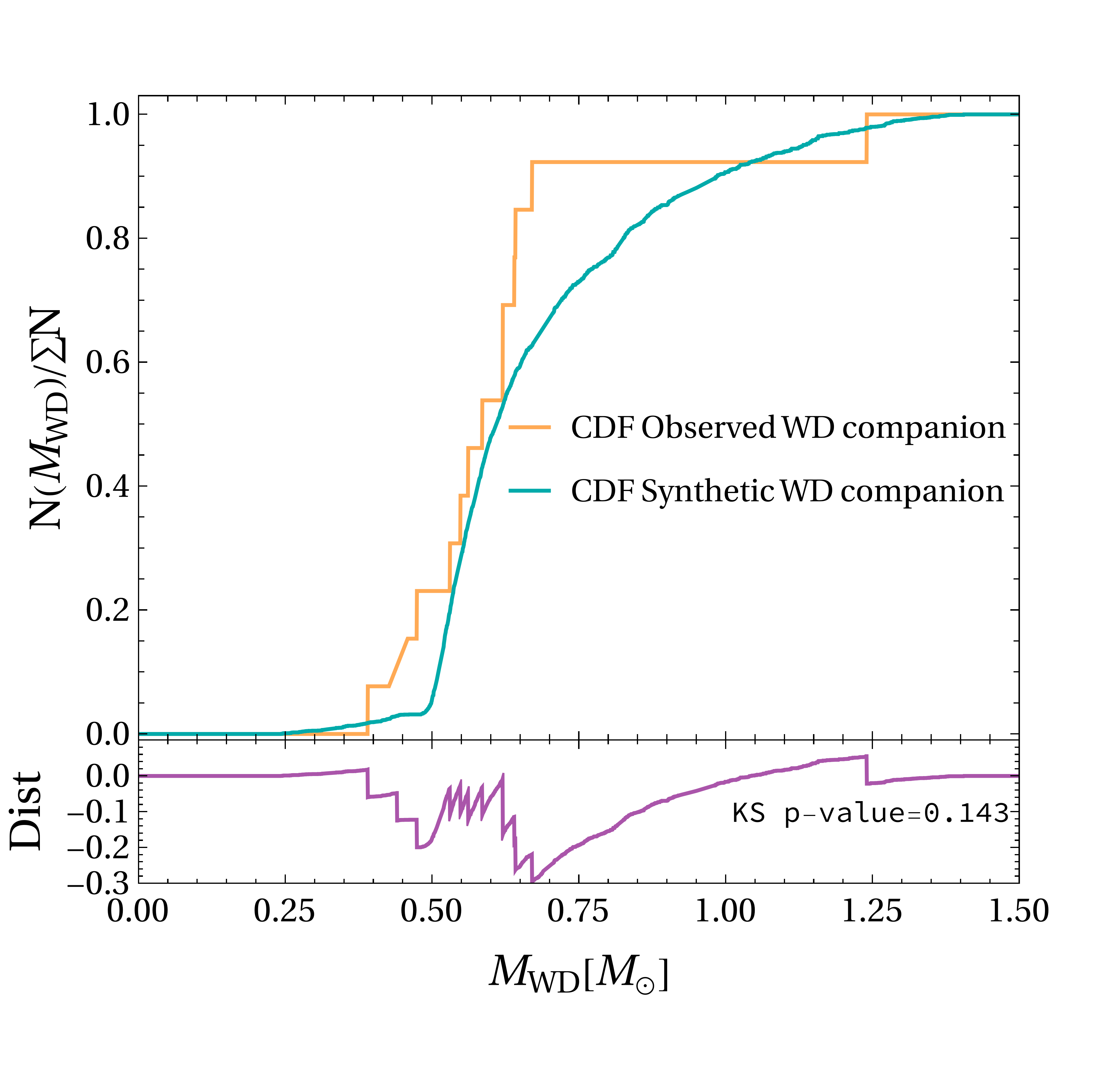}
\includegraphics[trim=0 0 0 0 clip=true, width=0.49\columnwidth]{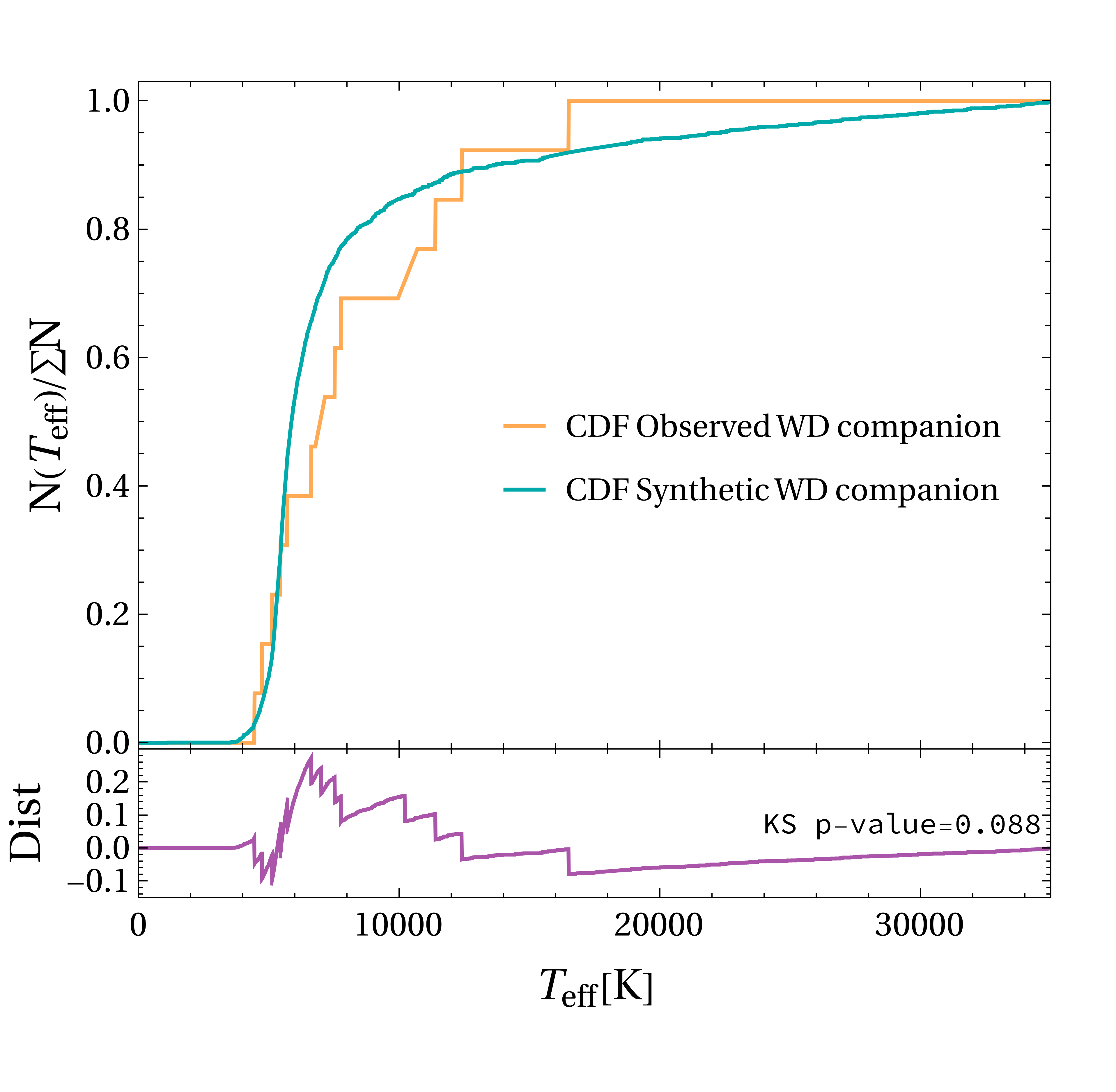}
\caption{ CDF of the synthetic sample (cyan line) and the observed sample at 100\,pc (orange line) for the mass (left panel) and the effective temperature (right panel) of the white dwarf companion after including BD models in our simulations. For a more quantitative comparison, the distance (DIST; magenta line) between the CDFs is also plotted.}
\label{f:KS_MTeff_WD}
\end{figure*}

\begin{figure*}[h!]
\centering
\includegraphics[trim=10 8 10 0 clip=true, width=0.485\columnwidth]{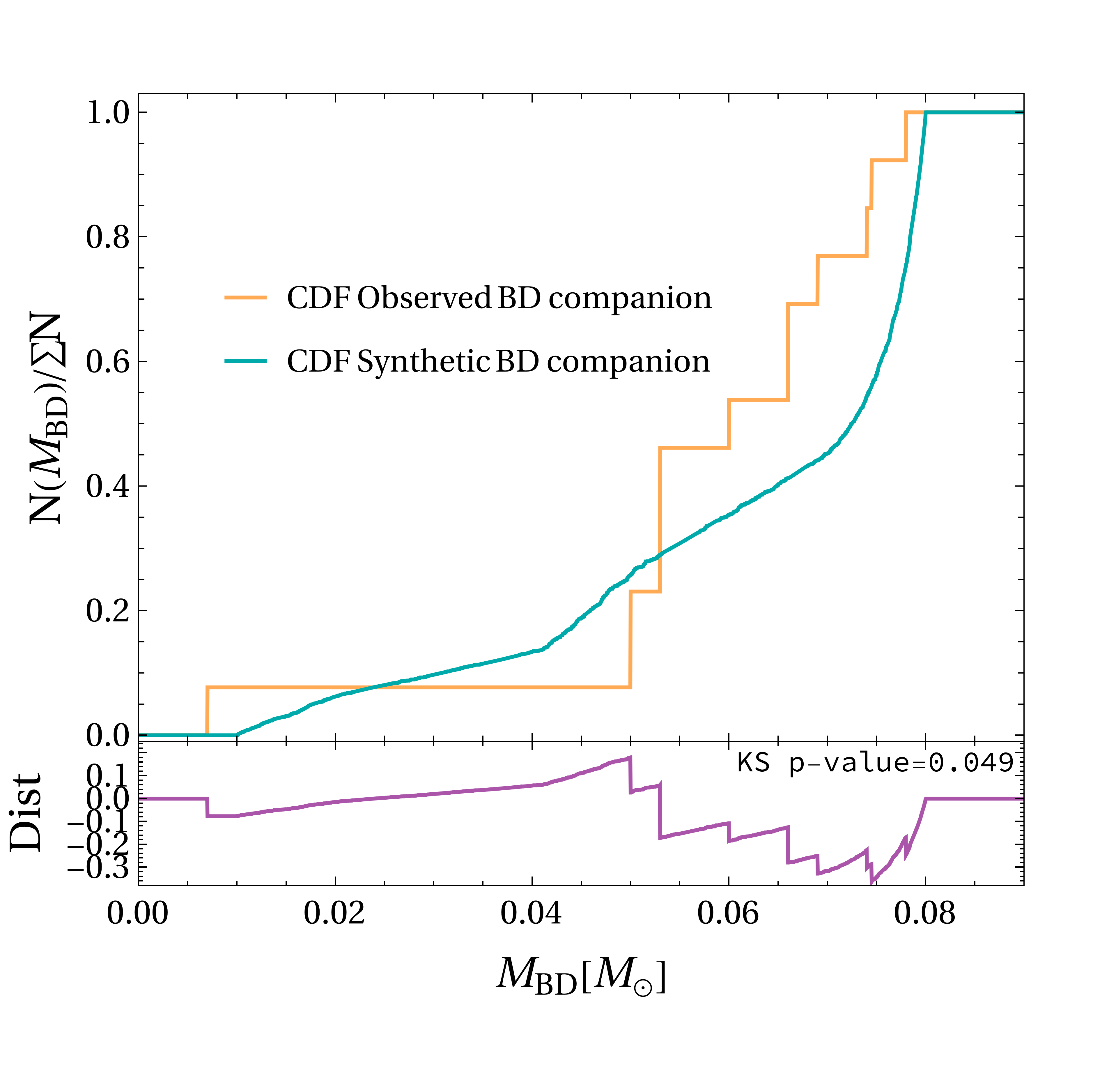}
\includegraphics[trim=0 0 0 0 clip=true, width=0.49\columnwidth]{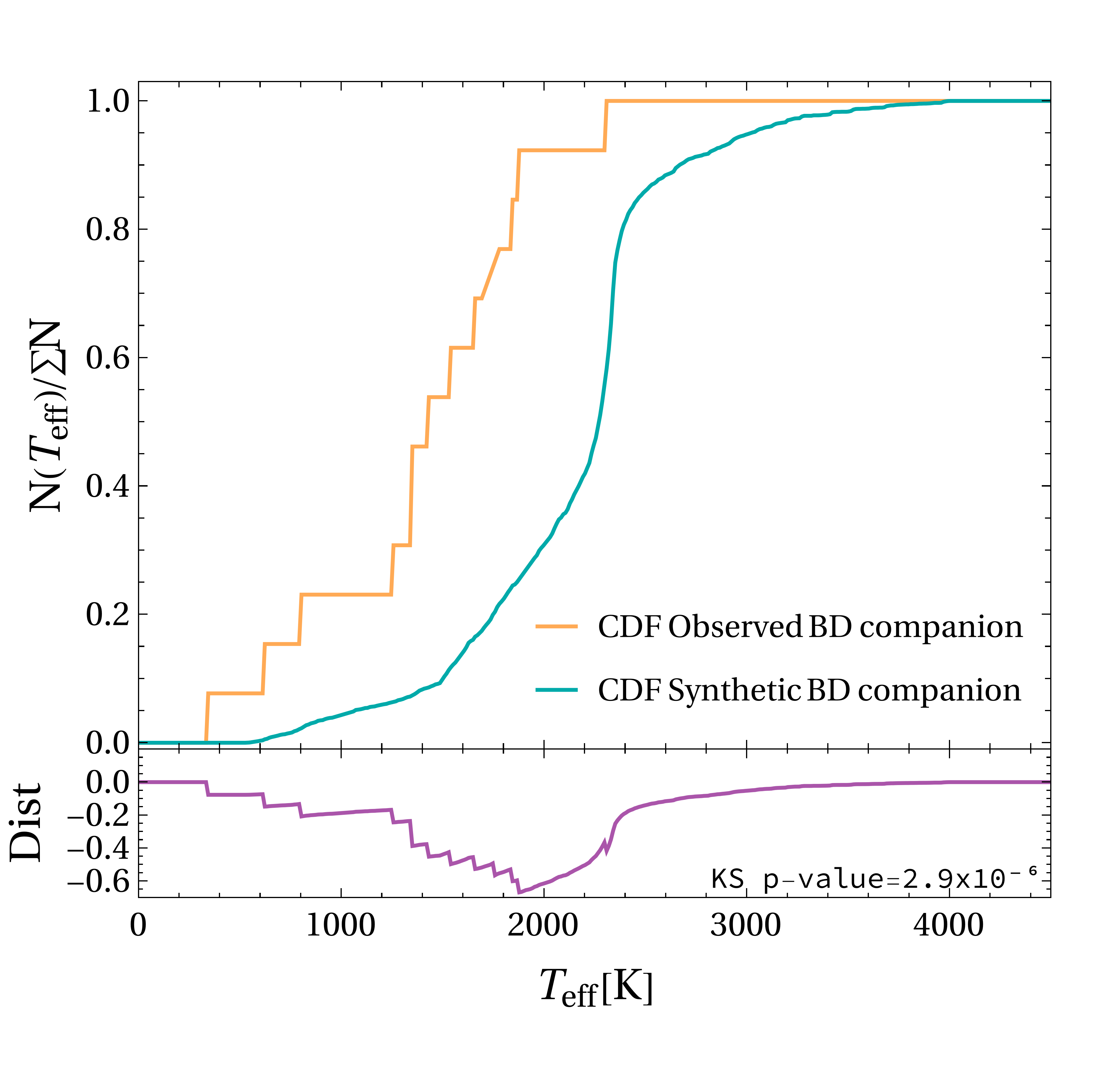}
\caption{ Same as in Fig. \ref{f:KS_MTeff_WD} but for the BD companion.}
\label{f:KS_MTeff_BD}
\end{figure*}

\begin{figure*}[h!]
\centering
\includegraphics[trim=0 0 0 0 clip=true, width=0.49\columnwidth]{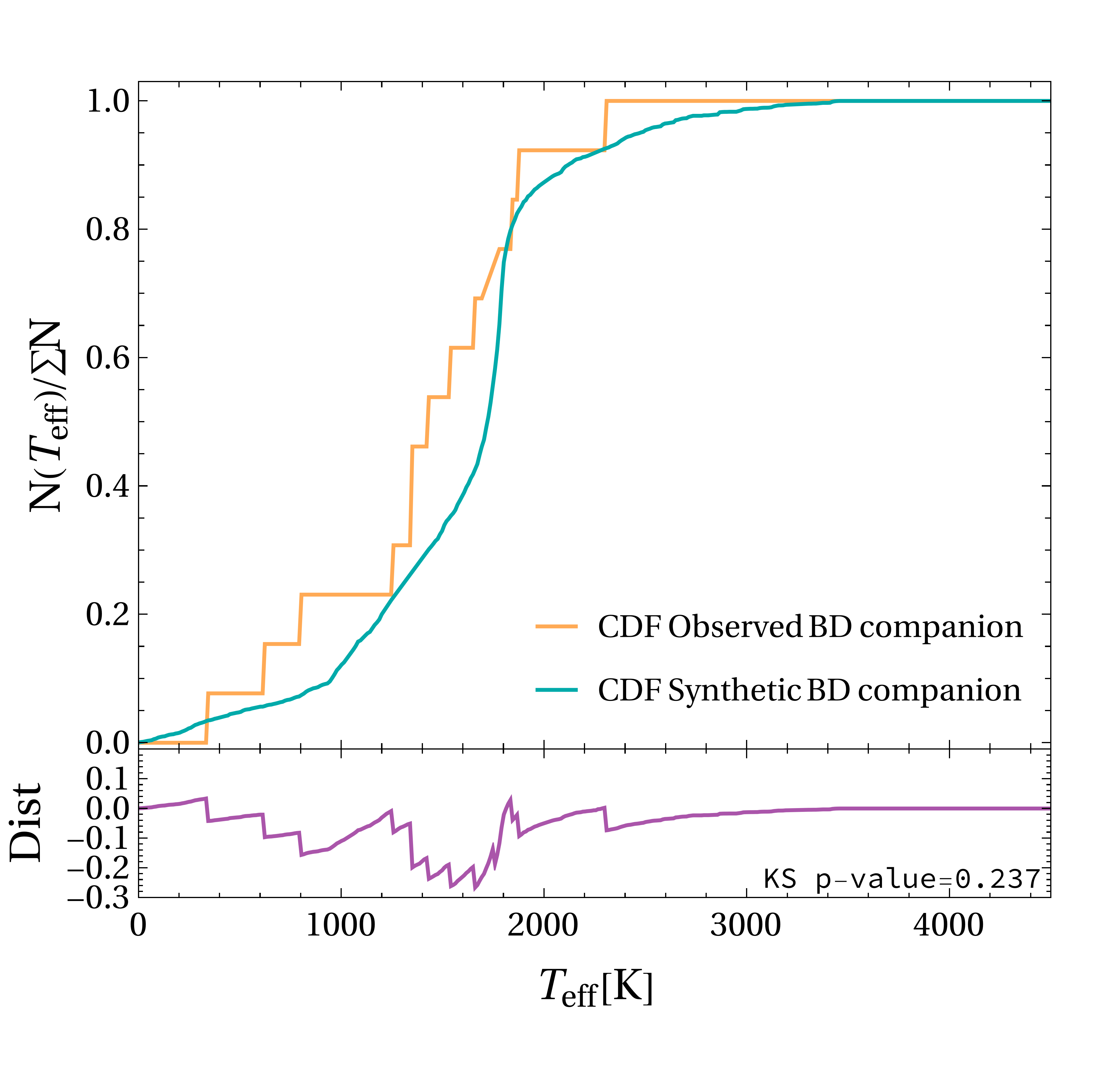}
\includegraphics[trim=0 0 0 0 clip=true, width=0.49\columnwidth]{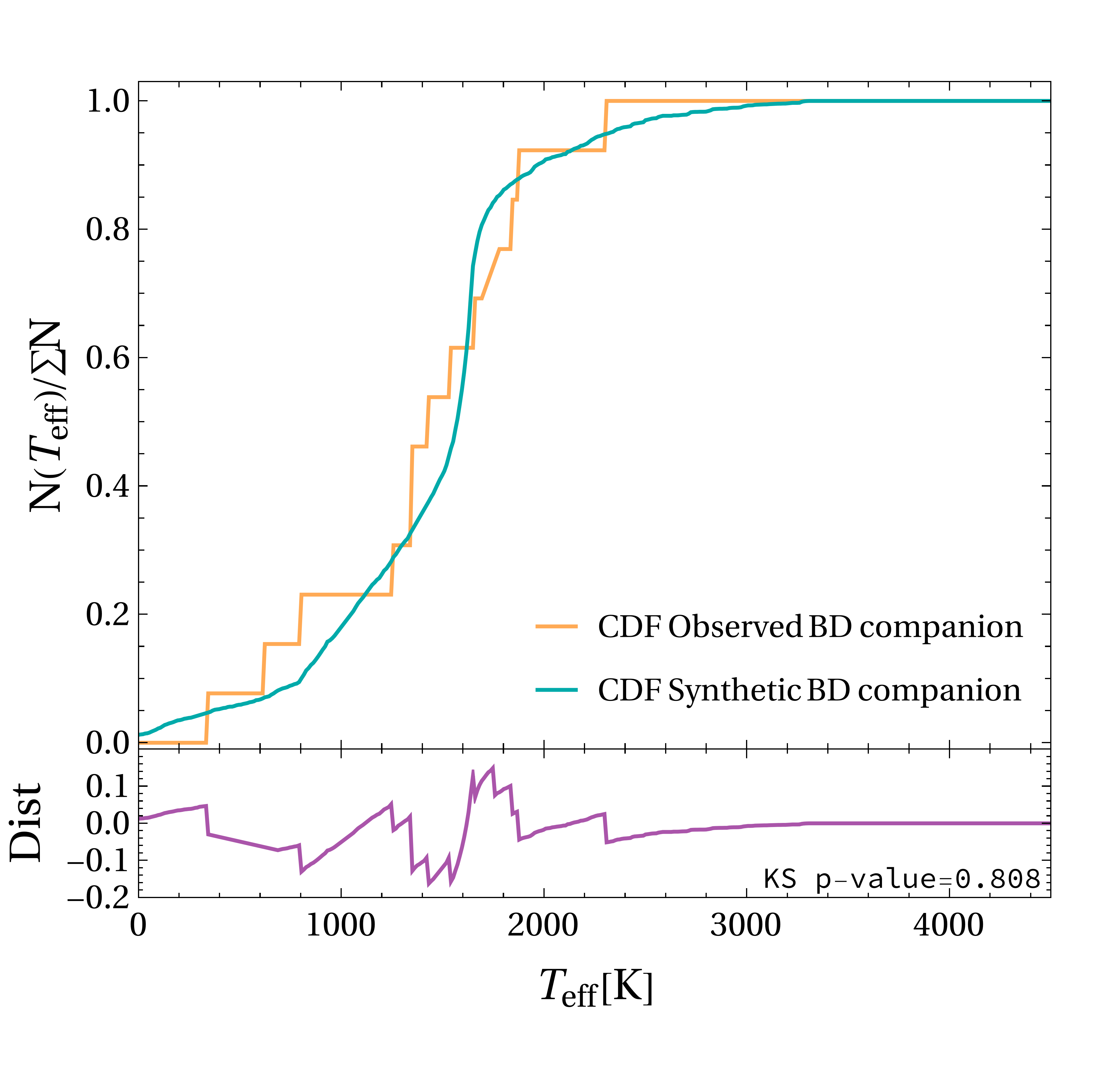}
\caption{ Same as in the right panel of Fig. \ref{f:KS_MTeff_BD} after sifting the models we used in 550K (left panel) and 700 K (right panel).}
\label{f:KS_Teff+700}
\end{figure*}

\begin{figure*}[h!]
\centering
\includegraphics[trim=0 0 0 0 clip=true, width=0.49\columnwidth]{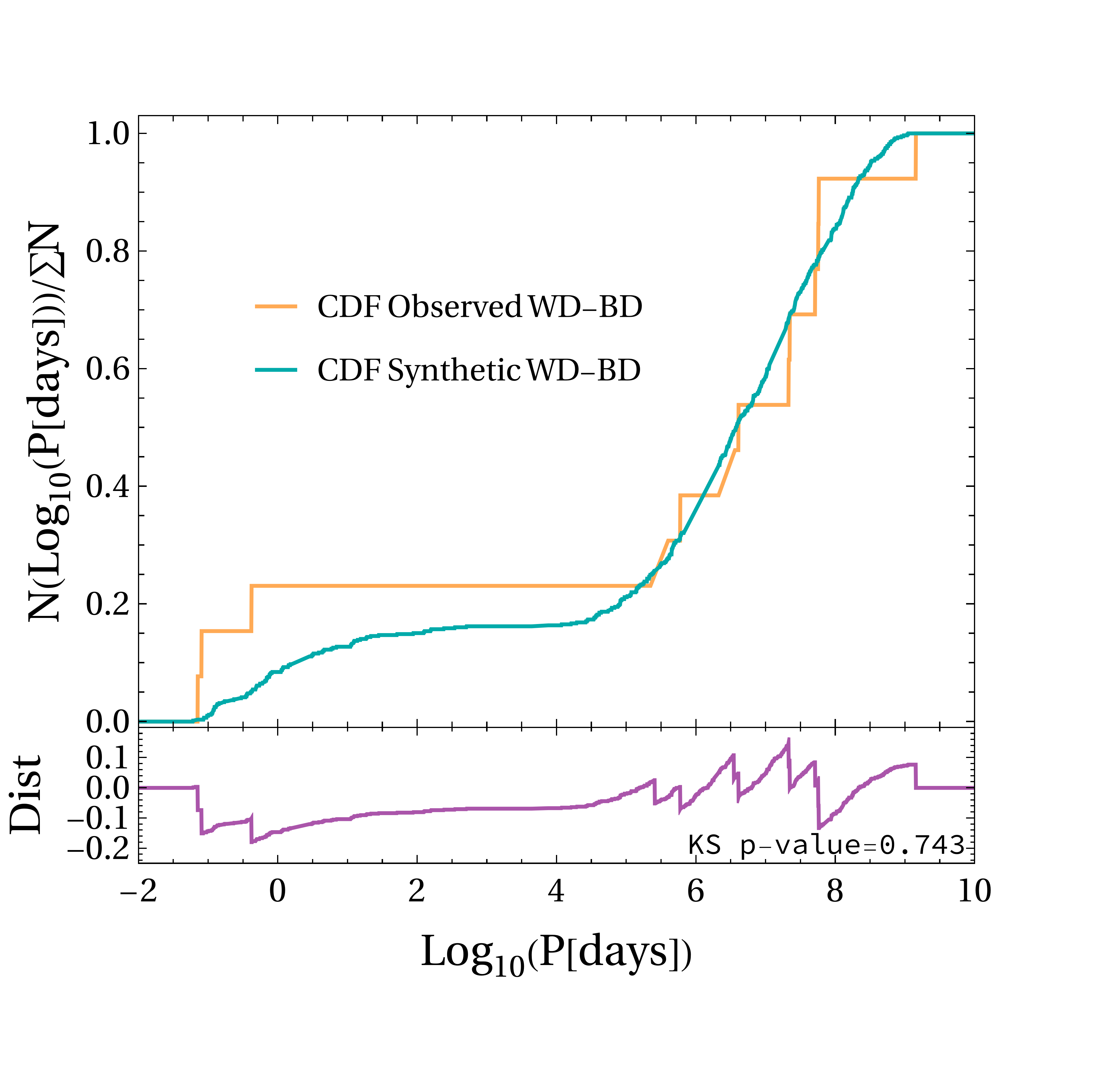}
\caption{ CDF of the synthetic sample (cyan line) and the observed sample at 100\,pc (orange line) for the period of the studied WD--BD systems after including BD models in our simulations. For a more quantitative comparison, the distance (DIST; magenta line) between the CDFs is also plotted.}
\label{f:KS_Per}
\end{figure*}

\begin{figure*}[h!]
\centering
\includegraphics[trim=0 0 0 0 clip=true, width=0.49\columnwidth]{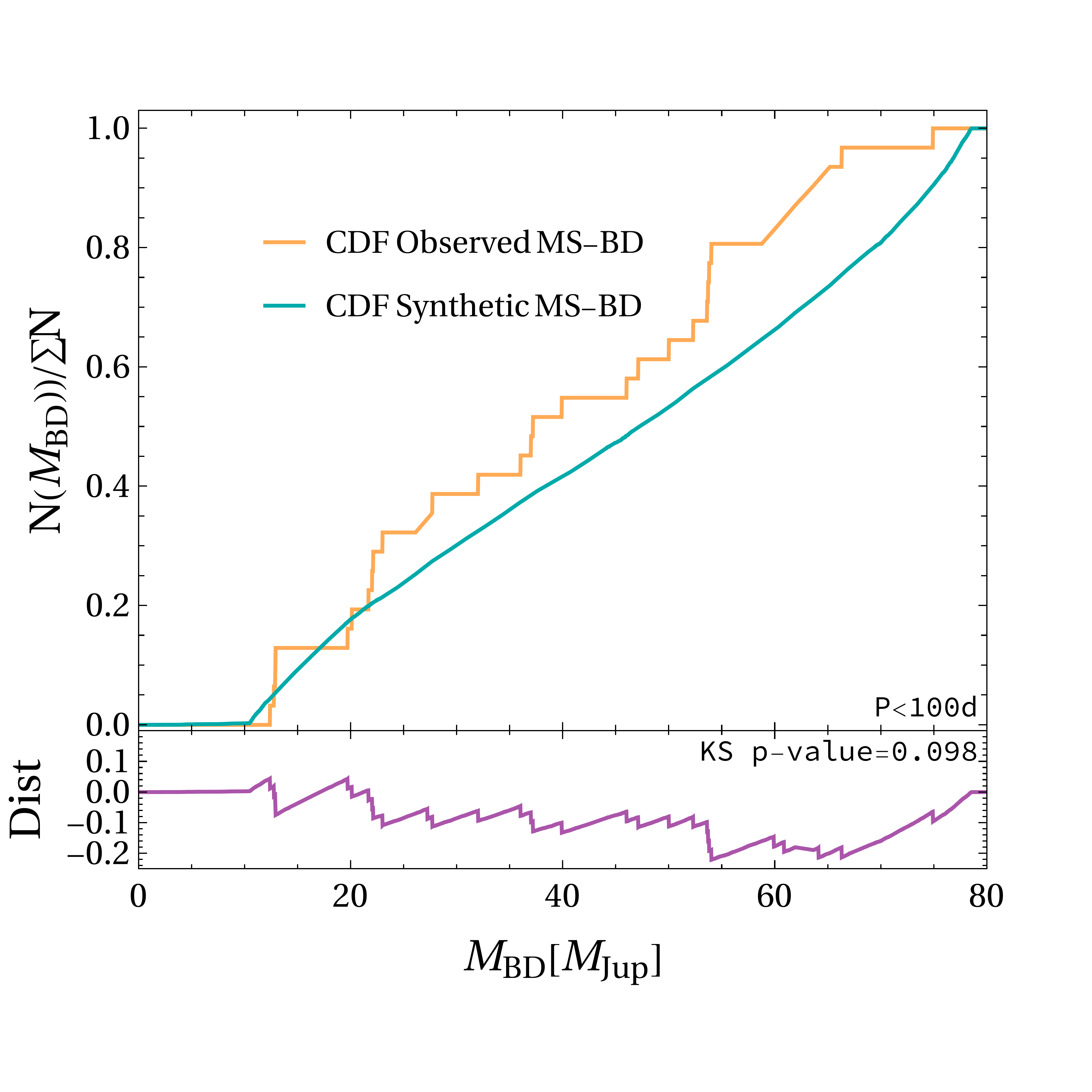}
\includegraphics[trim=0 0 0 0 clip=true, width=0.49\columnwidth]{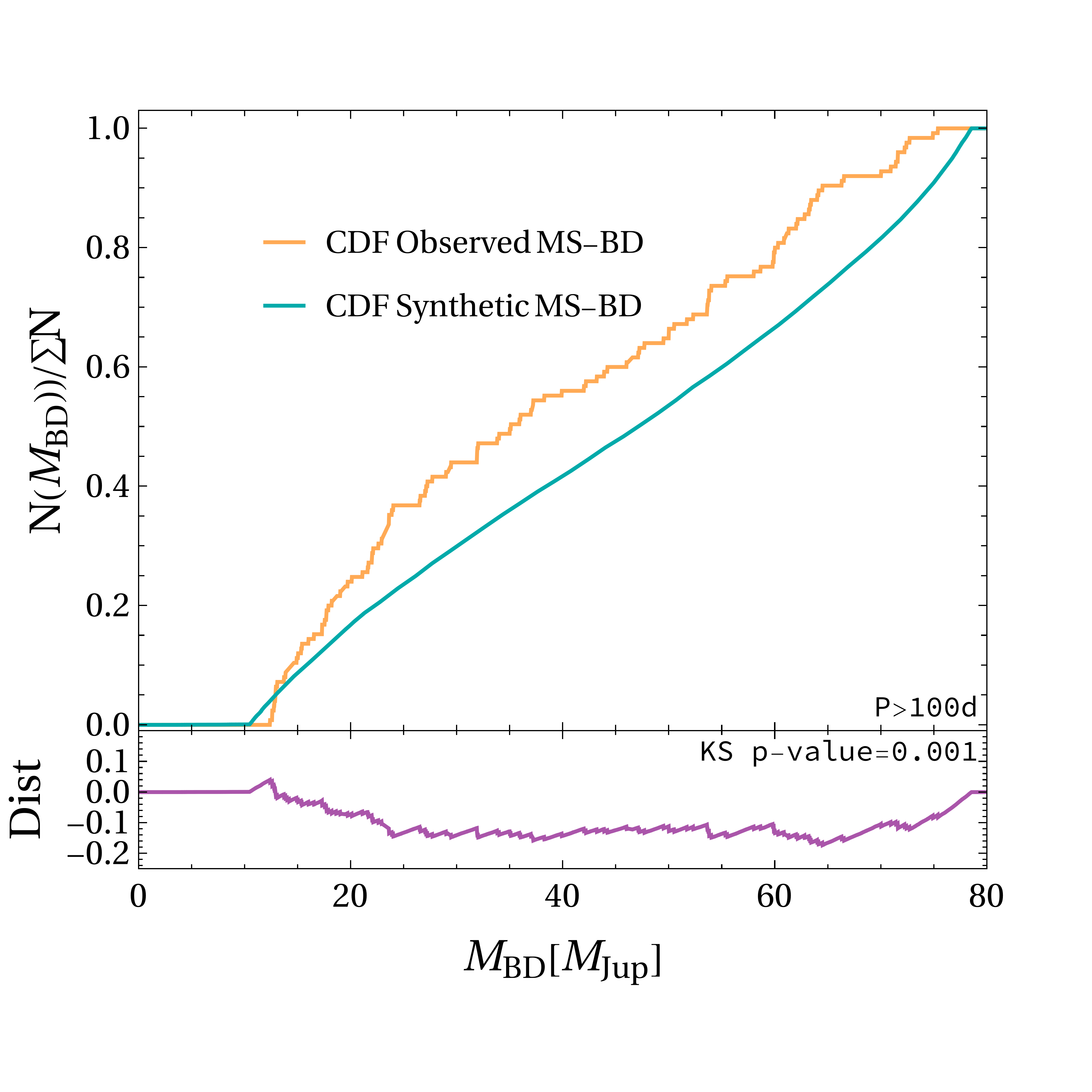}
\caption{ CDF of the synthetic sample (cyan line) and the observed sample at 100\,pc (orange line) for the mass of the brown dwarf companion in MS--BD systems with $Log_{10}P[days]$<100\,d (left panel) and $Log_{10}P[days]$>100\,d (right panel). For a more quantitative comparison, the distance (DIST; magenta line) between the CDFs is also plotted.}
\label{f:KS_M-BD}
\end{figure*}

\end{document}